# Removing Orbital Debris With Lasers*#

Claude R. Phipps*[1], Kevin L. Baker[3], Brian Bradford[5], E. Victor George[2], Stephen B. Libby[3], Duane A. Liedahl[3], Bogdan Marcovici[2], Scot S. Olivier[3], Lyn D. Pleasance[3], James P. Reilly[4] Alexander Rubenchik[3], David N. Strafford[5] and Michael T. Valley[6]

**Abstract**: Orbital debris in low Earth orbit (LEO) are now sufficiently dense that the use of LEO space is threatened by runaway collisional cascading. A problem predicted more than thirty years ago, the threat from debris larger than about 1cm demands serious attention. A promising proposed solution uses a high power pulsed laser system on the Earth to make plasma jets on the objects, slowing them slightly, and causing them to re-enter and burn up in the atmosphere. In this paper, we reassess this approach in light of recent advances in low-cost, light-weight modular design for large mirrors, calculations of laser-induced orbit changes and in design of repetitive, multi-kilojoule lasers, that build on inertial fusion research. These advances now suggest that laser orbital debris removal (LODR) is the most cost-effective way to mitigate the debris problem. No other solutions have been proposed that address the whole problem of large and small debris. A LODR system will have multiple uses beyond debris removal. International cooperation will be essential for building and operating such a system.

## Why Debris Clearing is Important

Thirty-five years of poor housekeeping in space have created several hundred thousand pieces of space debris larger than 1cm in the 400 -2000-km altitude low Earth orbit (LEO) band, their density reaching a peak in the 800-1,000-km altitude range[i]. Debris in the 1- 10-cm size range are most hazardous to LEO space vehicles because they are not tracked, but can cause fatal damage. For objects below 1 cm, "Whipple shields", though expensive, would be effective against hypervelocity impact[ii], and can be built. The range of debris orbit inclinations give a most probable closing velocity between objects[iii] of about 12km/s, a speed at which a piece of debris has ten times the energy density of dynamite. A 100-gram bolt would certainly cause a lethal event on the International Space Station, if it struck the crew chamber. Larger objects present a lesser threat, because they are less numerous (less than 10,000), and can be tracked and usually avoided by maneuvering. Even so, in March, 2009 and again on June, 2011, it was necessary for Space Station astronauts to take cover in a Soyuz capsule to reduce the chance of penetration by an

---

[1] Photonic Associates, LLC, Santa Fe, NM 87508 [* Member, AAAS; person to whom correspondence should be sent]

[2] Centech, Carlsbad CA 92011

[3] Lawrence Livermore National Laboratory, Livermore CA 94550*

[4] Northeast Science and Technology, Williamsburg, VA 23188

[5] ITT Space Systems Division, Vienna VA 22180

[6] Sensing and Imaging Technologies Department, Sandia National Laboratories, Albuquerque NM 87123**

* This work was performed in part under the auspices of the U.S. Department of Energy by Lawrence Livermore National Laboratory under Contract DE-AC52-07NA27344.

# Sandia is a multiprogram laboratory operated by Sandia Corporation, a Lockheed Martin Company, for the United States Department of Energy's National Nuclear Security Administration under contract DE-AC04-94AL85000.

object with unacceptable track uncertainty. Fortunately, the capsule was docked with the Station. Earlier, in February 2009, an American Iridium satellite collided with a Russian Kosmos satellite, and the resulting cloud of debris combined with that from the Chinese Fengyun 1C ASAT test in January, 2007, to greatly increase the density of debris around the Earth, prompting concerns about the safety of the final Hubble servicing mission. The instability predicted by Kessler and Cour-Palais[iv] has now reached the point where collisions are on track to become the most dominant debris-generating mechanism. While improved debris tracking and orbit prediction can temporarily improve threat avoidance via maneuvering[v], effective debris clearing strategies will eventually be necessary. Operational models of the changing risks of space debris damage have been developed to analyze costing strategies for debris removal[vi].

## Debris Threat Categories and Clearance Strategies

There are about $N_1 = 2{,}200$ large objects (diameter ≥ 100cm, mass of order 1 ton) in LEO, and $N_2 = 190$k small objects (diameter ≥ 1cm) [vii]. The flux for the small ones in the peak density region[viii] is about $R_2$ = 1.4E-4 $\mathrm{m^{-2}year^{-1}}$. Based on the relative numbers, we deduce a flux $R_1$ = 1.7E-6 $\mathrm{m^{-2}year^{-1}}$ for the large ones in the LEO band. Taking $\sigma = 2\mathrm{m}^2$ as the large object cross-section, the interval between collisions of type $i$ on the large ones across the ensemble is

$$T_{i1} = [\sigma N_1 R_i]^{-1} \quad . \tag{1}$$

Applying Eq. (1), the chance that a big object will impact a big object is once in $T_{11}$ = 134 years, whereas the chance a small object will impact a big object is once in $T_{21}$ = 3 years. Just removing the big derelicts does not solve the problem. Any new large space asset that is installed in LEO will encounter the same collision rate $R_{21}$ as before, from the small objects that have not been removed. The lifetime for these small objects at 1000 km altitude is of order 100 years[ix]. A system that can address the small objects as well as the big ones is needed.

Both classes need to be addressed because, while the debris growth rate is reduced by removing large derelict objects that produce clouds of debris when hit[x], the small-debris threat to a LEO asset is far larger numerically. For example, the chance of a fatal debris-caused Space Station event per decade is about 7%[xi]. Previously, removal of the small debris was underemphasized.

## Proposed Solutions to the Debris Problem

Aside from the laser-based approaches, including the pulsed laser ablation method that is the subject of this article, a variety of solutions have been proposed. To name a few, these have included chasing and grappling the object[xii], attaching deorbiting kits[xiii], deploying nets to capture objects[xiv,xv], attaching an electrodynamic tether[xvi,xvii,xviii] and deploying clouds of frozen mist[xix], gas[xx] or blocks of aerogel[xxi] in the debris path to slow the debris. While few of these concepts have progressed to the point where costs can be accurately estimated, Bonnal has estimated a cost of 27M$ per large object[13] for attaching deorbiting kits. Any mechanical solution will involve a comparable $\Delta v$, so we take Bonnal's estimate as representative of the removal cost per large item using mechanical methods.

The mist or unconfined gas solution would have effects that are not debris-specific. A mist or dust cloud deployed in LEO would rapidly disperse, as would a gas detonation, and, if sufficient mass were installed, it would cause existing space platforms as well as derelicts to re-enter.

The gas solution can avoid dispersal, but that requires the deployment of four hundred 100-km diameter balloons in orbit[20]. Even if they could be placed so as not to deny space to other assets, they are one-time solutions (one balloon per target) and costly to launch. If made of 5μm mylar, each 100km balloon would weigh 160 kilotons and cost $1,600 B to put in orbit using today's launch costs. [xxii],

The aerogel solution has similar problems. It is easy to show that[xxiii] an aerogel "catcher's mitt" solution designed to clear the debris in two years would require a slab 50cm thick and 13 km on a side. Such a slab would have 80-kiloton mass, and would cost $800M to launch. Even if we ignore the difficulty of maintaining this shape, a fatal problem is the steady 12kN average thrust required to oppose orbital decay of the slab facing ram pressure over an elliptical orbit ranging between 400km and 1100km altitude. To maintain this thrust over a two-year lifetime would require a fuel mass of 150 kilo-tons, in addition to the mitt mass, tripling the cost.

Laser-based methods can be divided into three general categories distinguished by their goals and laser beam parameters. At the lowest intensities, below the threshold for ablating the debris surface, lasers have been proposed to divert debris through the weak agency of photon momentum[xxiv]. This approach has laser momentum transfer efficiency four to five orders of magnitude less than pulsed laser ablation. It is problematic because its effects are comparable to the uncertain effects of space weather and sunlight, and does not effectively address the debris growth problem. At higher laser intensity, we can consider heating to ablation with continuous (CW) lasers, but slow heating of tumbling debris will usually give an ablation jet whose momentum contribution cancels itself out, on the average. CW heating causes messy melt ejection rather than clean jet formation, adding to the debris problem. Also, CW lasers cannot reach the required intensity on target at large range without a very small illumination spot size, requiring an unacceptably large mirror.

Pulsed laser orbital debris removal (LODR) was proposed fifteen years ago[3]. The basic setup is illustrated in Figure 1. At that time, lasers as well as telescopes with the required performance did not yet exist, but the components could be specified. Now, all the components actually exist or are in the planning stage.

As recently as four years ago[xxv], it was considered that "The use of ground based lasers to perturb the orbits of the satellites is not now practical because of the considerable mass of the satellites and the consequent need to deposit extremely high amounts of energy on the vehicles to effect the necessary change."

However, we believe that a better understanding of the problem, coupled with advances in technology driven by inertial fusion research, make this statement outdated. The purpose of this article is to demonstrate that laser orbital debris removal is practical and economical.

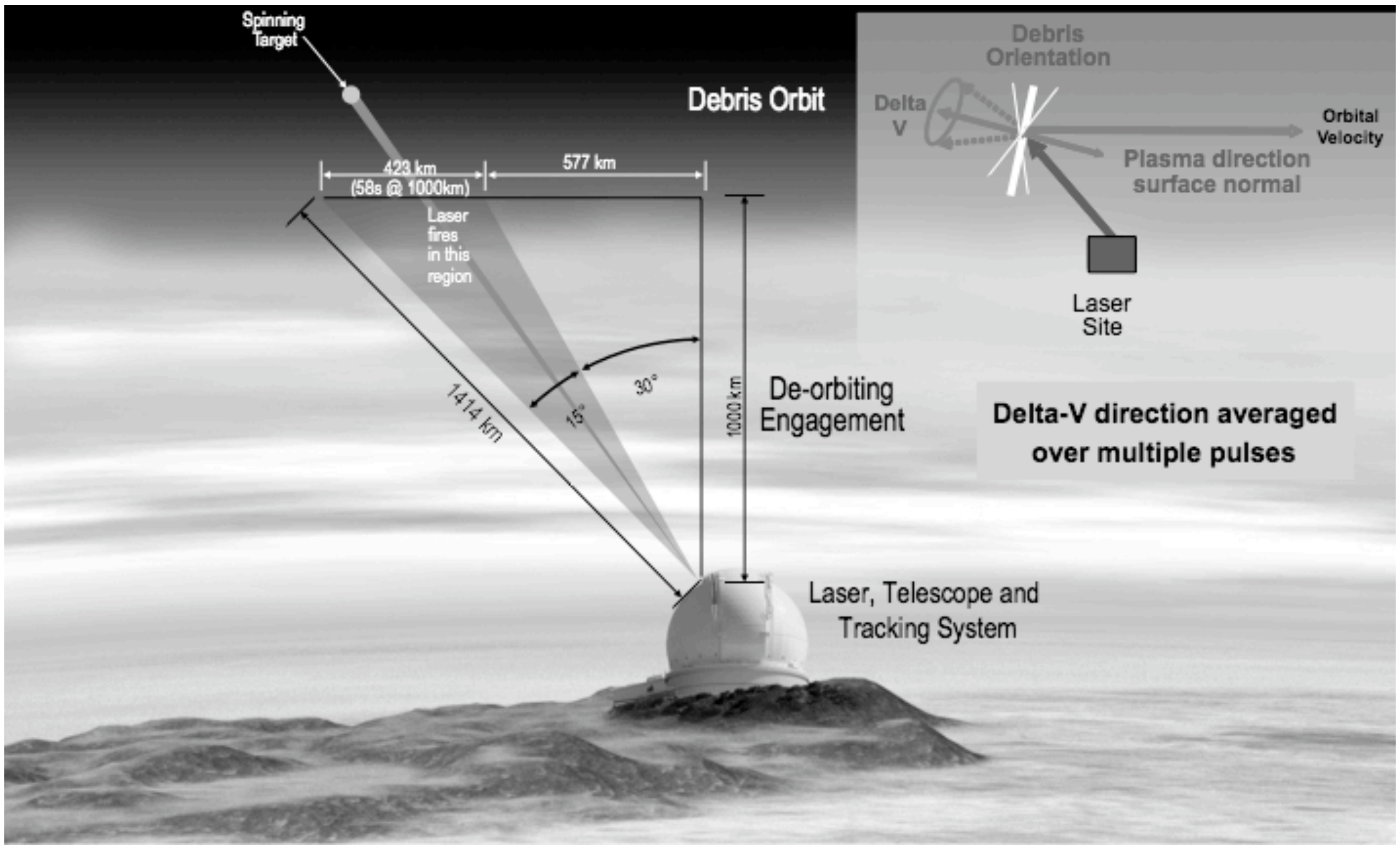

**Figure 1. LODR concept.**. The debris target is detected and tracked. Then, a repetitively pulsed laser is focused by a large mirror on it, making a plasma jet. With high intensity, 10 ns pulses, very little target material is removed and the debris is not melted or fragmented. Most of the laser energy goes into the jet. The engagement is designed so the jet points in the right direction to slow the target, on average, by the small amount (100-150 m/s) needed to drop its perigee to 200km, which is adequate for rapid re-entry**.** Hundreds of pulses are needed to do this, but they can be applied during one pass overhead for the small debris.

## How Lasers can Transfer Momentum

The standard measure of the efficiency with which laser light is converted to pressure is the momentum coupling coefficient,

$$C_m = p/I \text{ [Pa W}^{-1}\text{m}^{-2} \text{ or N/W].} \qquad (2)$$

In the ablation regime, $C_m$ is a function of the laser intensity $I$, wavelength $\lambda$ and laser pulse duration $\tau$ and material properties. As the intensity increases, $C_m$ rises to maximum and decreases at higher laser intensity, because more energy goes into reradiation, ionization, breaking chemical bonds, etc. Figure 2 shows[xxvi] this classical behavior. The maximum momentum coupling occurs just at the vapor-plasma transition. In order to design a LODR system, it is crucial to predict the fluence (laser energy per m$^2$) where this maximum is found, and this requires knowing how to combine[xxvii] vapor and plasma models for a particular material. An approximate working relationship is given by[27, xxviii,xxix]

$$\Phi_{opt} = 4.8\text{E}8 \sqrt{\tau} \text{ Jm}^{-2} \qquad (3)$$

For $\tau$ = 5ns, precise calculations show $\Phi_{opt}$ = 53 kJ/m$^2$ required for an aluminum target[27], a worst case.

Multi-kJ laser pulse energy and large mirrors are required to overcome diffraction spreading of the light at a range of 1000km. The spot size $d_s$ which can be delivered to a target at distance z is

$$d_s D_{eff} = aM^2\lambda z. \tag{4}$$

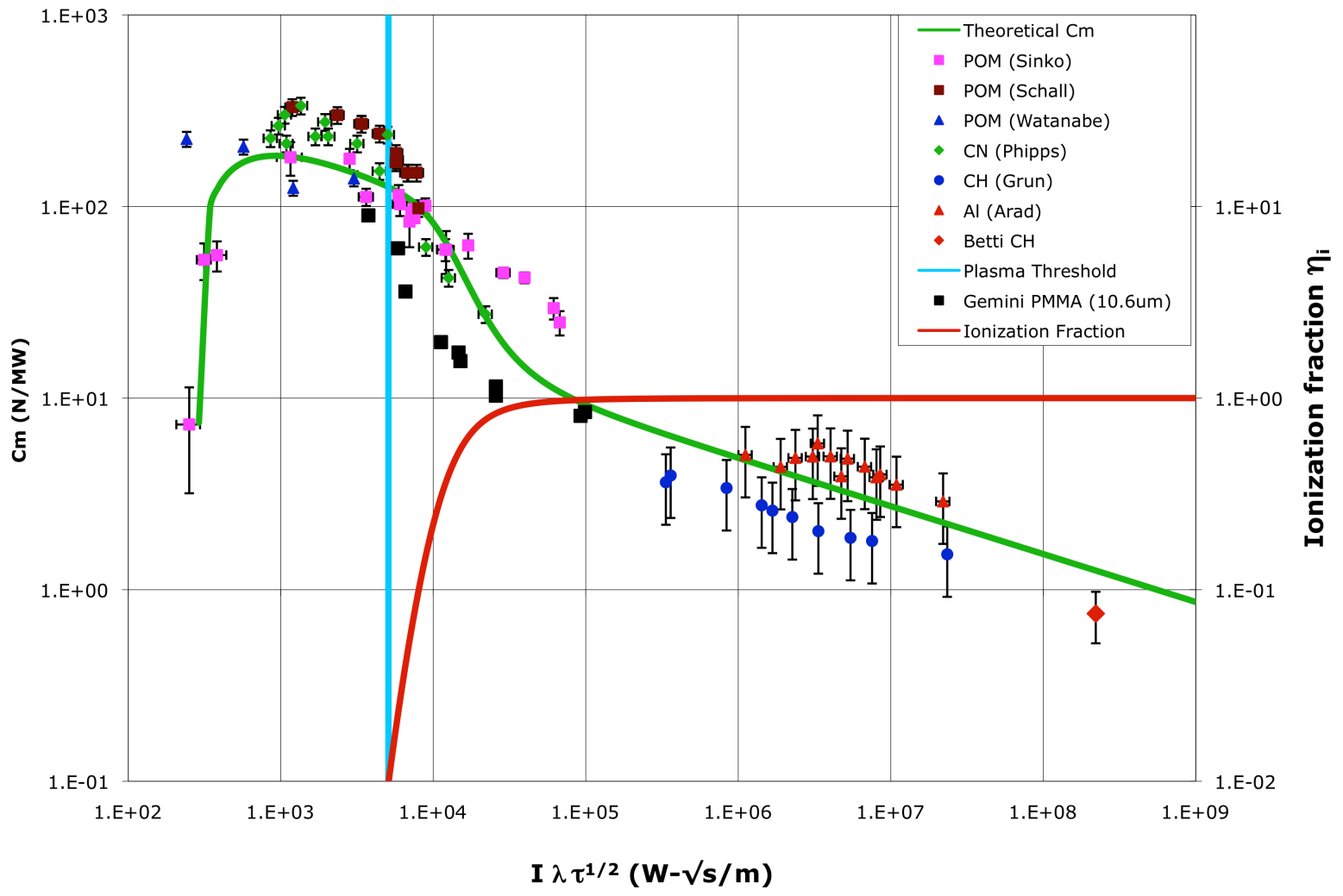

**Figure 2. Example results** of models that allow us to predict $C_m$ for a number of likely plastics and metals. The vertical blue line shows where the vapor-plasma transition implied by Eq. (2) occurs for $CO_2$ lasers, but the $C_m$ model is universal and applicable to a wide range of laser parameters. The red line is ionization fraction. References for the data are found in reference 23.

In Eq. (4), $M^2$ is the beam quality factor ($\geq$1)and $D_{eff}$ is the illuminated beam diameter inside the aperture $D$ for calculating diffraction. A hypergaussian[xxx] with index 6 coming from a LODR system with corrected beam quality $M^2$=2.0 (Strehl ratio = 0.25) gives $D_{eff}/D$ = 0.9 and $a$ = 1.7. In order to obtain even $d_s$ = 31 cm at z = 1000km range with $\lambda$=1.06μm we need $D_{eff}$ = 13m illuminated aperture diameter and, to avoid nonlinear effects in the atmosphere, a minimum $D_{eff}$ = 11m. The quantity $M^2$ in Eq. (4) includes atmospheric phase distortions corrections, either by standard adaptive optics or phase conjugation or a combination of the two (discussed below and in the SOM).

Lightweight mirrors of this size[xxxi,xxxii] now have a major impact on LODR system design. Examples are the 10-m Keck primary, the 9.8 x 11.1-m South African Large Telescope, and the planned European Extremely Large Telescope with a 42-m diameter primary mirror composed of 984 segments[xxxiii] with very low areal mass density.[32,xxxiv]

Denoting by $T_{eff}$ the product of all transmission losses, including apodization, physical obscuration by the secondary mirror, spider, coudé path and atmospheric transmission loss, Eq. (4) shows that the fluence $\Phi$ delivered to a target by laser pulse energy $W$ is

$$\Phi = \frac{4WT_{eff}D_{eff}^2}{\pi M^4 a^2 \lambda^2 z^2}. \tag{5}$$

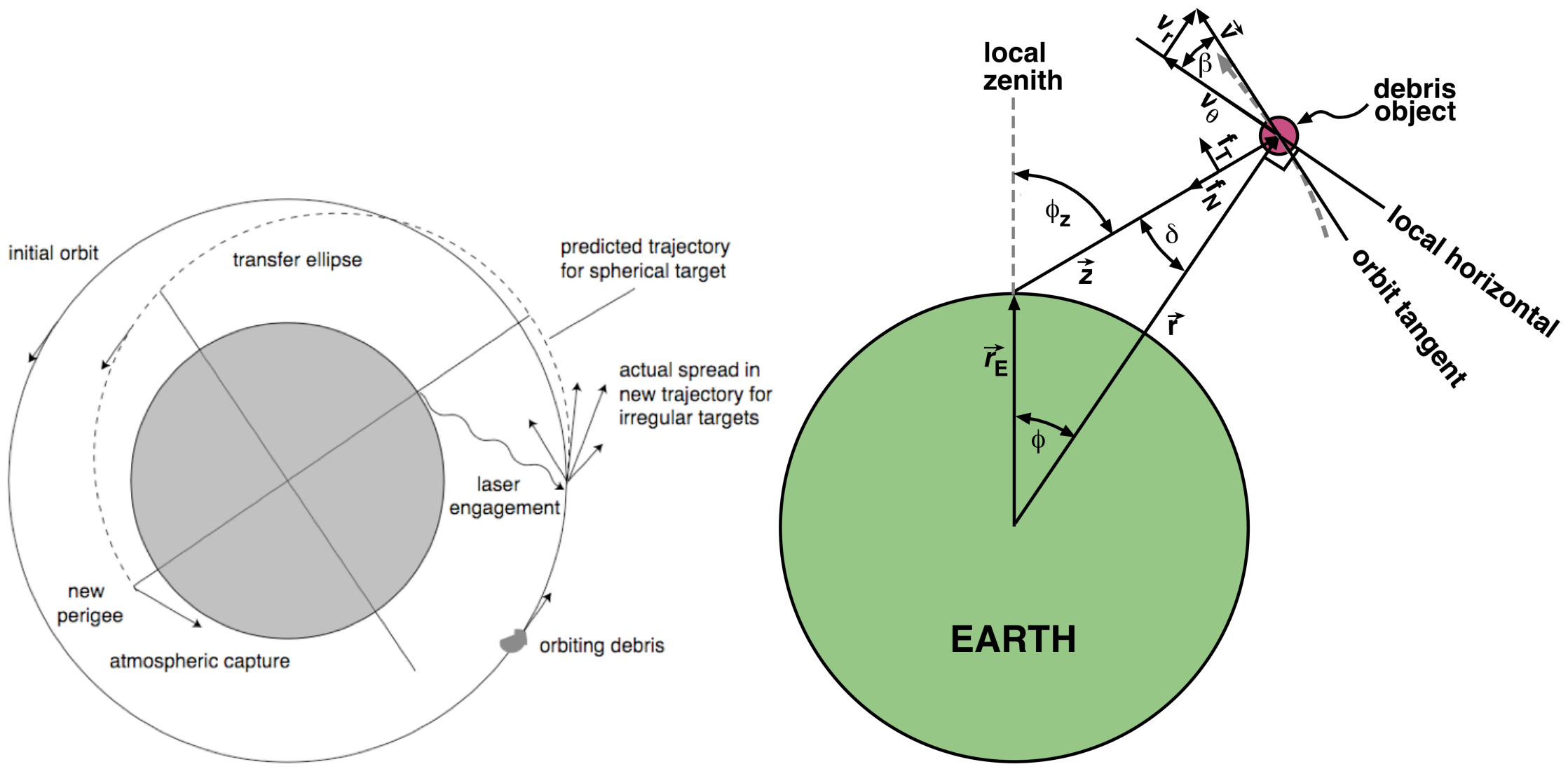

3a. 3b.

**Figure 3. Geometry of the laser-target interaction** (scales exaggerated for clarity).
**3a:** Schematic of debris de-orbiting concept in low-Earth orbit. For a given energy deposition, the orbital perturbation on a spherical target is predictable. For non-spherical targets, the perturbation can be predicted, if the shape and orientation at engagement are known.
**3b:** Thrust on a debris object is resolved into components $f_T$ and $f_N$ normal to and along the orbit tangent. Since, for LEO debris, range $z$ << the Earth's radius $r_E$, the zenith angle $\phi_z$ changes rapidly compared to the geocentric angle $\phi$.

In a practical case where $D_{eff}$ = 10m, if $T$ = 80%, $T_{eff}$ = 0.5. In order to deliver 53 kJ/m$^2$ to a target at 1000km range, the product $WD_{eff}{}^2$ must be at least 993 kJm$^2$, laser pulse energy must be 7.3kJ, and the mirror diameter $D$ must be 13m.

Predicting the velocity increment $\Delta\bar{v}$ delivered to a debris object is not a simple matter. It depends on target shape and the orientation of each surface element. The thrust from the plasma jet formed on the target is perpendicular to each element, whatever the angle at which the laser strikes the target. Further, the engagement has to be properly designed to make sure $\Delta\bar{v}$ is slowing the target rather than speeding it up.

To simplify discussion, we use an efficiency factor $\eta_c$ for the combined effects of improper thrust direction, target shape, target tumbling, etc. in reducing the efficiency of laser pulse fluence $\Phi$ on the target (J/m$^2$) in producing the desired velocity change,

$$\Delta v_{||} = \eta_c C_m \Phi/\mu. \qquad (6)$$

In Eq. (6), $\mu$ is the target areal mass density (kg/m$^2$). This formulation takes account of laser beam "overspill" for small debris, without having to specify the actual size and mass of each target. We take $\eta_c$ = 0.3 after Liedahl[xxxv] [see the SOM for a complete discussion of target shape effects]. In some cases, we can lower the debris perigee not only by pushing antiparallel to its velocity vector, but, counterintuitively, by pushing radially outwards.

We take $|\Delta v_o| =$ 150m/s needed for LEO re-entry and $\mu$ = 10kg/m$^2$ for a small target. This value of $\mu$ is an upper bound average value for small debris.[xxxvi] $C_m$ can range from 50 to 320 μN-s/J just for various surface conditions of aluminum.[xxxvii] We have shown $C_m$ values up to 300 μN-s/J for various organics representative of space debris[26]. For illustration we use $C_m$ = 75μN-s/J. With these values, Eq. (6) shows we have $\Delta v_{||}$ = 12cm/s for each laser shot. Taking target availability to be T=100s, repetition frequency for the 7.3 kJ laser pulse must be $(\Delta v_o/\Delta v_{||})/T$ = 12.5Hz, for a time-average laser power of 91kW. If the target were as big as the beam focus, it would have 0.75kg mass. Smaller targets of whatever mass with this mass density would also be caused to re-enter in a single overhead pass, even though the beam spills around them.

## Re-entry of Small Targets

Figures (5a) and (5b) show calculations for targets up to 1 kg mass and range up to 1000km being de-orbited in a single overhead pass. Apsidal rotation occurs, but is irrelevant for single-pass re-entry, since the target does not have to be re-acquired. With apogee overhead, only 100 seconds illumination are needed for re-entry. We averaged over the possible orbital orientations to obtain the Table 1 results. The Table 1 system may be considered a "starter system."

## Re-entry of Large Targets

It has been claimed that lasers cannot de-orbit large, 1-ton derelict debris objects that are of concern. Indeed, single-pass re-entry of these objects is not possible. However, large debris are catalogued and have reasonably accurate ephemerii. Let's consider a 1-ton target with area A = 1.25m$^2$ presented to the laser (Table 2). With the parameters listed in the Table, it takes 3.7 years to re-enter one object. However, 167 different objects can be addressed in one day, giving 4.9 years to re-enter the whole constellation. Note that it is only necessary[xxxviii] to re-enter 15 of these large objects annually to *stabilize* the debris environment. From this standpoint alone, the LODR system is a good investment. A larger mirror is required for the large-target system to avoid nonlinear effects in the atmosphere.

**Table 1. Small-target LODR System Parameters**

| Target Parameters | | Optical System Parameters | |
|---|---|---|---|
| Maximum mass (kg) | 0.75 | Wavelength λ (μm) | 1.06 |
| Areal Mass Density μ (kg/m$^2$) | 10 | Pulse Length τ (ns) | 5 |
| Maximum Range (km) | 1,000 | $C_m$ (μN-s/J) | 75 |
| Perigee Altitude (km) | 500 | Active mirror diameter $D_b$ (m) | 13 |
| Apogee Altitude (km) | 700 | Spot Size on Target (m) | 0.31 |
| Useful Apparition (s) | 200 | Fluence on Target (kJ/m$^2$) | 75 |
| Minimum Permitted Elevation (°) | 30 | Pulse Energy (kJ) | 7.3 |
| Retargeting Time (min) | 1.0 | Repetition Frequency (Hz) | 11.2 |
| System Availability (%) | 50 | Average Optical Power (kW) | 81 |
| Number of Targets Accessible | 100k | Push Efficiency $\eta_c$ | 0.30 |
| Time to Re-enter all Targets (mo) | 8.7 | Average Interaction Duration (s) | 100 |
| | | Beam Quality Factor | 2.0 |
| | | Beam Hypergaussian Index | 6 |

## Multiple Uses

LODR systems would be useful for purposes other than complete re-entry of all large debris, such as:

Increasing ephemeris precision: building a LODR system necessitates detection and tracking technology that permits location of targets with 1m precision, far better than present practice. This capability will allow more accurate collision prediction.

Orbit modification on demand for large objects: Even the small-target LODR system would then be able to nudge these objects to avoid collisions, or to provide modest orbit changes, inducing as much as a 35 cm/s velocity change in a 1,000 kg target during a single overhead pass. This is more than required to divert a large target and avoid a predicted collision.

| Table 2. Large-target LODR System Parameters | | | |
|---|---|---|---|
| **Target Parameters** | | **Optical System Parameters** | |
| Mass (kg) | 1,000 | Wavelength $\lambda$ ($\mu$m) | 1.06 |
| Areal Mass Density $\mu$ (kg/m$^2$) | 820 | Pulse Length $\tau$ (ns) | 10 |
| Maximum Range (km) | 1,500 | $C_m$ ($\mu$N-s/J) | 75 |
| Perigee Altitude (km) | 500 | Mirror diameter $D_b$ (m) | 25 |
| Apogee Altitude (km) | 900 | Target Spot Size [defocused] (m) | 1.25 |
| Useful Apparition (s) | 250 | Fluence on Target (kJ/m$^2$) | 75 |
| Apparition Interval (days) | 10 | Pulse Energy (kJ) | 140 |
| Minimum Permitted Elevation (°) | 60 | Repetition Frequency (Hz) | 2.7 |
| Retargeting Time (min) | 1.0 | Average Optical Power (kW) | 370 |
| System Availability (%) | 50 | Push Efficiency $\eta_c$ | 0.30 |
| Number of Interactions for Re-entry | 135 | Average Interaction Duration (s) | 250 |
| Time to Re-enter one Target (yrs) | 3.7 | Beam Quality Factor | 2.0 |
| Targets Addressed Per Day | 167 | Beam Hypergaussian Index | 6 |
| Number of Targets | 2,200 | | |
| Time to Re-enter all Targets (yrs) | 4.9 | | |
| Effective Re-entry Rate per year | 450 | | |

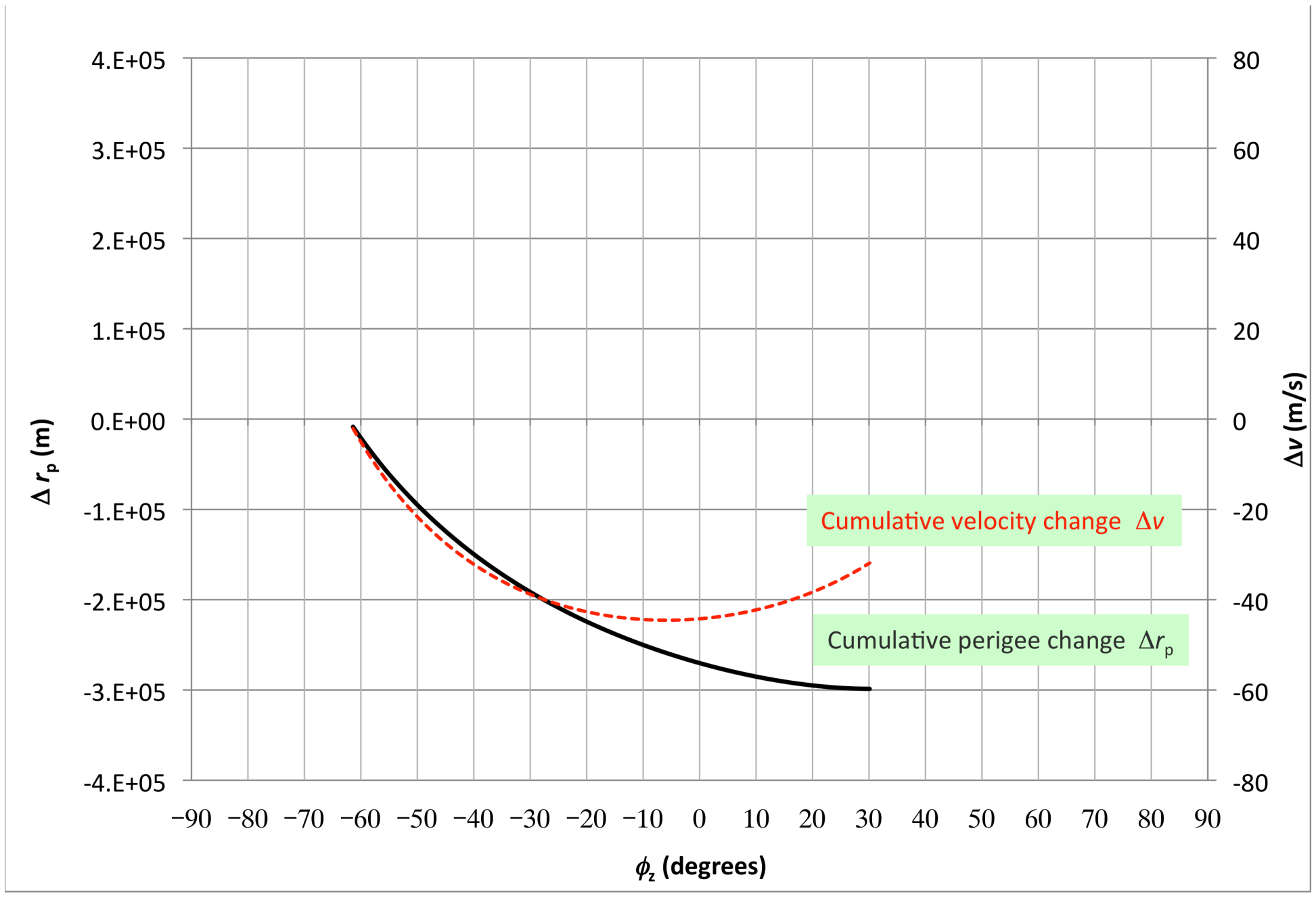

5a.

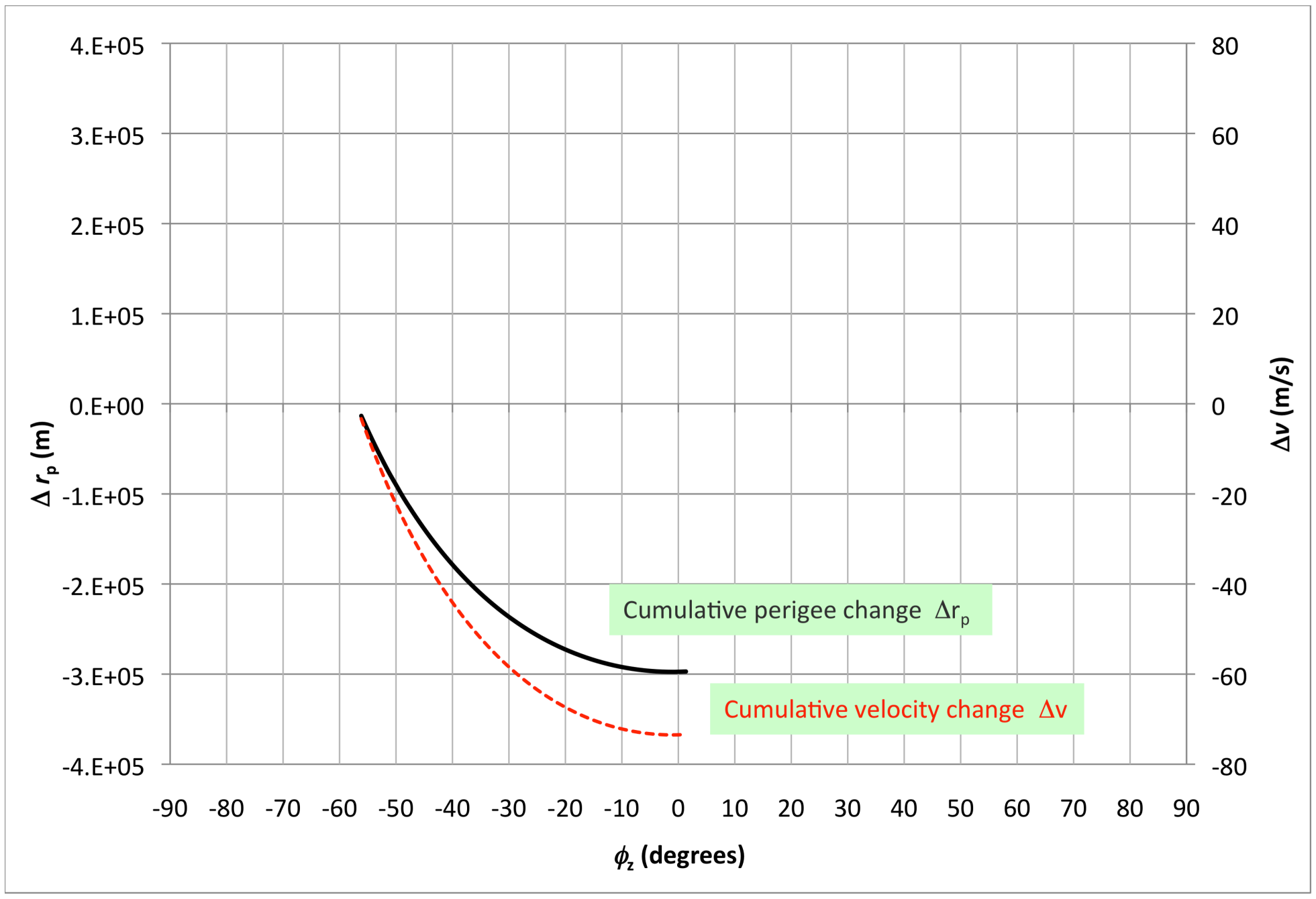

**5b.**

**Figure 5. Target re-entry** is achieved in one overhead pass for <u>any</u> debris target smaller than the laser spot radius of 31cm at 1000 km range, having areal mass density 10kg/m$^2$ or less. The largest target re-entered has 0.75kg mass. Parameters: Wavelength 1.06 μm, beam quality factor 2.0, beam format hypergaussian with index 6, fluence on target 53 kJ/m$^2$, 7.3kJ pulse energy, repetition rate 11.2 Hz, mirror diameter 13 m, $C_m$ = 75 μN-s/J, efficiency factor $\eta_c$ = 30%, perigee altitude 500km, apogee altitude 1073km, eccentricity 0.04, re-entry for $\Delta r_p$ = -3E5m.
Case a): orbit perigee is -120 degrees geocentric (upstream) relative to laser site, 833 pulses applied all along the debris path over 210 s to achieve minimum perigee. Case b): apogee is overhead the laser site, 1,010 pulses applied over 133 s.

Causing precise re-entry: Re-entry for selected derelicts can be altered in a calibrated fashion so the re-entry site can be controlled, for example, in the mid-Pacific, avoiding liability issues as well as damage to property or people on the ground.

Moving GEO targets into disposal orbits: The small target system, coupled with a 10-20m relay mirror just above geosynchronous (GEO) orbit is capable of raising the orbit of a defunct GEO satellite 100km in just 20 minutes.

## Acquisition

An acquisition system reduces the position uncertainty of a debris object from km to the meters required by the laser system. An easy method would use twilight or dawn, when the target is in sunlight and the acquisition system sees dark sky, but this would limit operation to about two hours per day.

Active acquisition is possible, in daylight[9], using the "pusher laser" to illuminate the target, and the LODR system mirror on Earth to collect the scattered light. The laser beam is first deliberately defocused to cover several km at range. If the field of view is 3km at 1000km range, one object per 4 minutes will pass through the field of view on average.. A large (20m) receiving aperture and 7.3kJ pulses from the pusher laser are required to provide enough scattered photons to see small targets. InGaAs focal plane arrays now have quantum efficiencies of 80%[xxxix]. In our active tracking system, a 1.5-cm Lambertian scattering target with 50% albedo at 750 km range would return 45 photons to its array pixel on the ground, with a signal to day sky background ratio of 72. The system would require a bandwidth of 0.2nm for both the laser and narrowband optical filter, and a 75 km "range gate."

The 20-m mirror has two parts with different optical quality. The central 13m section used by the laser in pusher mode is high quality. The 3.5-m annulus outside that is used to collect light for initial wide field of view acquisition and can be lower quality, since we need only a few-m image precision in the target plane in this mode. If we have a 1000x1000 element array with a 3-km field of view, each pixel projects onto a 3-m spot. Both parts use segments about 1m in size mounted on three-point piezoelectric mounts. The outer annulus can be pointed at a different spot from the central portion. Four independent adaptive optics systems are required [see "Phase Correction" following and SOM Figure 4].

The optical filter is easy to obtain. Range gating amounts to reading out the array every 250μs and storing the data in slices, delayed from laser firing by the propagation time. This gives rough range information.

## Target Tracking

When the acquisition system has established a track within a 3-km circle, the field of view is narrowed, always keeping the target centered. When the circle viewed has shrunk to 100m, the system switches modes [see "Look-ahead" following].

As the field of view is narrowed, the focal plane array is protected from damage with attenuators. Now, the computer makes the best foci possible and the pusher laser begins doing its real work [see "Phase Correction" following].

## Phase Correction via Adaptive Optics

Phase aberrations are caused by several mechanisms, from thermal distortions in the laser amplifier to atmospheric turbulence. The conventional solution is adaptive optics[xl], in which a deformable phase plate with many computer-driven actuators compensates for these distortions as they occur, at a rate of about 1kHz (SOM). The phase reference for such a system is provided by a “guidestar.” Examples are a 100W beam at 589nm that creates a starlike reference point source in the Earth’s sodium layer at 90km altitude, and the reflection from the target itself.

## Look-ahead

The finite velocity of light requires dealing with “look-ahead” before an accurately tracked target can be “pushed.” At 7.5km/s, the debris is actually as much as 50 m ahead of where the sensor last detected it. Range information is needed to tell the computer how much to correct pointing for the pusher shot, because the target’s actual speed and distance are critical variables. The tracking system outlined above can do this. The laser now appears to be shooting into empty space but, when its pulse arrives, the target is there.

We are literally looking in two directions at once, separated by about 100μrad. Two independent adaptive optics systems correct these paths. After the mode change, the detector path continues using the target itself as guidestar. Meanwhile, a sodium laser guidestar is tilted ahead of the detector by a computed angle, and a separate array uses the signal from that to command the corrector plate that helps the laser focus on its target.

The fine tracking signal now becomes very bright and shifts into the blue as plasma is formed on the target. The system uses this signal to stop increasing laser pulse energy.

## Rôle of Brillouin-Enhanced Four-Wave Mixing (BEFWM) in Adaptive Optics

BEFWM (see Figure 7 and SOM) is a type of phase conjugation in which phase distortions are automatically compensated[xli,xlii,xliii] This is important when mirror size becomes much larger than atmospheric turbulence cells, because conventional adaptive optics require thousands of actuators operating at a 1kHz rate.

It may be easier to use BEFWM than classical adaptive optics, or perhaps a hybrid system will be best. Phase conjugation operates like holography, but it is a dynamic hologram dynamically recorded by interfering waves in a nonlinear optical medium rather than being a static pattern on a glass plate. With a phase conjugate mirror, each ray is reflected back through the system in the direction it came from with reversed phase. This reflected wave "undoes" the distortion, converging to the initial point source. The amplified conjugate signal is automatically concentrated on the space object to an accuracy that is determined not by the turbulent scattering angle (~100 μrad) but, instead by the spatial resolution of the receiving aperture (~ 0.1 μrad for a receiving aperture of 10 m).

A special advantage of this technique is that the target becomes its own guidestar.

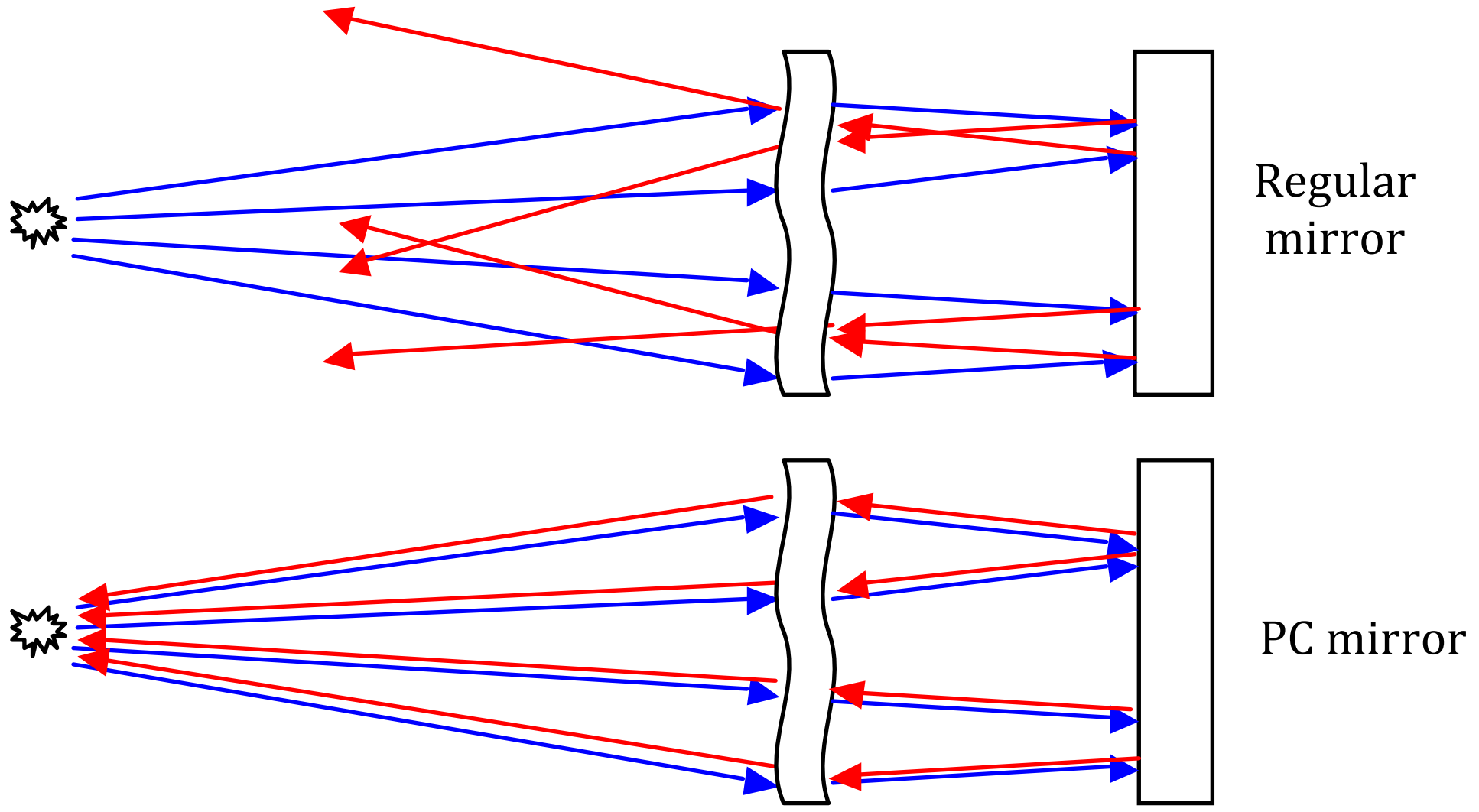

**6a.**

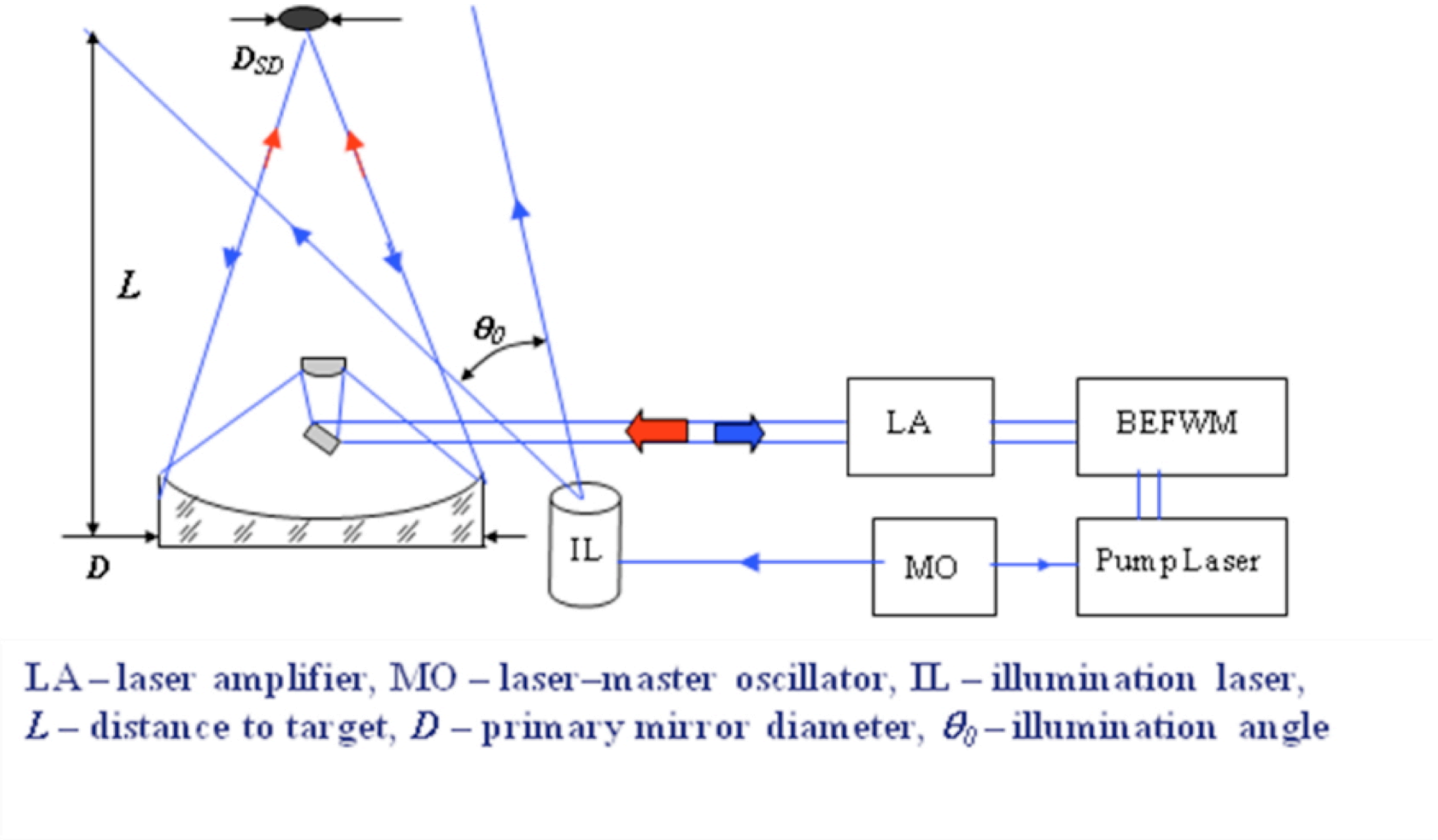

**6b.**

**Figure 6. How BEFWM works.** a): A phase conjugate (PC) mirror behaves differently from a regular mirror. A regular mirror reflects incident rays in the opposite direction, so that the angle of incidence is equal to the angle of reflection (see Fig.1). In contrast, light from a phase conjugate mirror is always reflected exactly in the backward direction, independent of the angle of incidence. b): A nonlinear optical cell (BEFWM) and pump laser are added to the usual laser chain to implement a BEFWM system.

Other advantages are that tilt anisoplanatism is eliminated, and that the system has extremely narrow acceptance bandwidth leading to good background noise rejection. The time by which the phase correction is "out of date" is just that required for a double pass through the atmosphere (~100μs), much faster than the 1ms time in which atmospheric phase distortions can typically change. Target lead-ahead is computed by a proprietary technique.

## Advances in Lasers

Laser systems built and operated at Lawrence Livermore National Laboratory (LLNL) over the past decade encompass the range of energies and powers required to remove orbital debris.

One example is the solid-state heat capacity laser (SSHCL), which was built and operated in the mid-2000's. This flush lamp-pumped, solid-state laser operated in burst-mode for a period of 10 s, produced 500 J pulses and average power of > 10 kW. [44]

Since 2009 LLNL has operated the world's largest and most energetic laser, the National Ignition Facility (NIF).[45,46] Combined NIF's 192 laser beam produce over 3 MJ in 5-10 ns pulses at the fundamental wavelength (1053 um), and over 1.5 MJ at the third harmonic (353 nm). Building on what has been learned and demonstrated on the NIF, LLNL is now developing designs for a laser driver for the Laser Inertial Fusion Energy (LIFE) program.[47] This high-repetition rate (10-20 Hz), high-efficiency (~12-18%) diode-pumped solid-state system will produce 8-10 kJ in a single beam at 1053 nm.

LODR requires significantly less than 1% of the NIF pulse energy, does not require harmonic conversion and does not have the laser fusion energy driver requirement to operate 24/7 with high availability. A LODR laser will be simpler, more compact and far less costly than either the NIF or the LIFE laser system,[48] but will leverage the experience gained and investment made over several decades of laser development, construction and operation.

## Demonstration System

A demonstration system could be built using a 9-m mirror and a 4.6-kJ laser to prove LODR works on targets at 400km altitude.

## International Cooperation

Building and operating a LODR system will require international cooperation to avoid concerns that it is really a weapons system. Also, cooperation in its operation will be needed to facilitate permission for its use to remove large debris objects.

## Estimated Debris Removal Cost

We do not claim high accuracy for our cost models. An accurate model requires a thorough engineering study. However, rough system cost estimates based on the algorithms described in the Project ORION review[9] are useful to estimate cost per object re-entered. We estimate cost per small object removed at a few thousand dollars, and that for large objects at about $1M each.

## Conclusions

We analyzed all the major aspects of laser orbital debris removal, and conclude that laser orbital debris removal will work, even for large debris objects. A LODR system should provide the lowest cost per object removed among all approaches that have been proposed. LODR is the only solution that can deal with both small and large debris. With LODR, target access is at the speed of light, redundant and agile. LODR can handle tumbling objects, while mechanical grapplers cannot. The system has multiple uses aside from general debris clearing, such as preventing collisions, increasing the accuracy of debris ephemerii and controlling where large debris impact the Earth's surface. Development and construction of the laser debris removal system offers the opportunity for international cooperation. Indeed, such cooperation will be be necessary to avoid concerns that it is a weapon system and provide a framework for its practical use.

# Supporting Online Material

## Contents

Figures

## Introduction

This material provides additional details supporting the claims made in the main paper "Removing Orbital Debris With Lasers" by Phipps, *et al.* We review the physics of laser momentum coupling to targets, laser orbit modification using this coupling and the constraints on the ODR beam parameters posed by propagation through the atmosphere. We also review target shape effects, acquisition and tracking, atmospheric turbulence correction, the Brillouin-enhanced four-wave mixing technique as a possible alternative to standard adaptive optics and the methods for choosing targets.

## Laser Momentum Coupling

In the plasma regime, it has been shown[xliv] that the relationship

$$C_m = 1.84E-4 \frac{\Psi^{9/16} / A^{1/8}}{(I \lambda \sqrt{\tau})^{1/4}} \quad \text{N/W} \qquad (1)$$

describes $C_m$ to within a factor of two for surface absorbers in the plasma-dominated regime and pulses longer than about 100ps. Also,

$$I_{sp} = 442 \frac{A^{1/8}}{\Psi^{9/16}} (I \lambda \sqrt{\tau})^{1/4} \quad \text{s} \qquad (2)$$

for the plume "specific impulse," $v_{plume}/g_o$. In Eqs. (1 and 2),

$$\Psi = \frac{A}{2[Z^2(Z+1)]^{1/3}} , \qquad (3)$$

where A is the average atomic mass number. The quantity $Z \geq 1$ is the average ionization state in the laser-produced plasma plume, and is also a function of ($I$, $\lambda$, $\tau$) because of its dependence on electron temperature in the plasma plume,

$$kT_e = 0.256 \frac{A^{1/8} Z^{3/4}}{(Z+1)^{5/8}} (I \lambda \sqrt{\tau})^{1/2} \quad \text{[eV].} \qquad (4)$$

The approximate value of Z is determined by applying Saha's equation[xlv],

$$\frac{n_e n_j}{n_{j-1}} = \frac{2u_j}{u_{j-1}} \left( \frac{2\pi A m_p k T_e}{h^2} \right)^{3/2} \exp\left(-W_{j,j-1}/kT_e\right) \qquad (5)$$

and writing

$$Z = n_e/n_i, \qquad (6)$$

with

$$\sum_{j=1}^{j\,max} (n_j) = n_i \quad . \qquad (7)$$

Parameters in the preceding relationships are: $W_{j,\ j-1}$, the ionization energy difference between the (j-1)th and jth ionization states of the material; $m_e$, the electron mass; Planck's constant $h$; $c$, the speed of light; $I$ the incident laser intensity (W m$^{-2}$); the plume electron total number density $n_e$ (cm$^{-3}$); $u_j$ the quantum-mechanical partition functions of the jth state; and $n_j$, the number density of each of the ionized states.

Predicting $C_{mv}$ in the vapor regime is more complicated and two models are used, depending on the data available for a particular material. For polymers in the vapor regime for which an ablation threshold fluence $\Phi$ (J/m$^2$) has been measured, we have shown[xlvi]

$$C_{mv} = \sigma / \Phi = \sqrt{\frac{2\rho C^2(\xi - 1)\ln\xi}{\alpha \Phi_o \xi^2}} \tag{8}$$

$$I_{spv} = \sqrt{\frac{2\alpha\Phi_o(\xi - 1)}{\rho g_o^2 \ln\xi}} \tag{9}$$

For elemental materials such as aluminum for which tables of vapor pressure vs. temperature $p(T)$ exist, e.g., the Los Alamos SESAME tables or Lawrence Livermore's QEOS or PURGATORIO-based equation of state models[xlvii],, we can work backward from hydrodynamic variables based on wavelength-independent material parameters to the incident intensity $I$ which must exist to balance these variables, obtaining[xlviii]

$$I = \frac{pv}{a}\left(\frac{\gamma}{\gamma-1}\right)\left[1 - \frac{T_o}{T} + \frac{q}{C_p T} + \frac{\gamma-1}{2}\right] + \frac{\sigma\varepsilon}{a}T^4 + B(\tau) \tag{10}$$

where

$$B(\tau) = \frac{1}{a}\left[\phi(T, x_h) + \frac{x_h \rho_s C_v (T - T_o)}{\tau}\right] \quad . \tag{11}$$

These expressions can be used to generate a numerical solution which relates ablation pressure $p$ and vapor velocity $v$ to $I$ over a range corresponding to $p(T)$ data, and we can compute the vapor regime coupling coefficient (for elemental materials such as aluminum) and specific impulse from

$$C_{mv} = p/I \quad . \tag{12}$$

$$I_{sp\ v} = v/g_o \quad . \tag{13}$$

In any case, we can model the transition[3] between the vapor and plasma regimes by writing for the combined coupling coefficient,

$$C_m = p/I = [(1-\eta_i)p_v + \eta_i p_p]/I = (1-\eta_i)\ C_{mv} + \eta_i\ C_{mp} \tag{13}$$

where the ionization fraction (the proportion of ionized to total plume particles including the neutrals $n_o$)

$$\eta_i = n_i/(n_o + n_i) \quad (14)$$

$\eta_i$ is determined numerically by iterating the process indicated in Eqs. (5-7). It is convenient to implement this iteration numerically (see Allen[xlix]) by forming

$$S_j = \frac{n_{ij}}{n_{i,j-1}} = \frac{8.64E26}{n_e} \frac{2\, u_j}{\theta^{1.5} u_{j-1}} \exp[-W_{j,j-1}/kT_e] \quad (15)$$

where $\theta = 5040/T_e$, and then computing the array

$$P_j = \prod_{k=1}^{j} S_k = \left[\frac{n_1}{n_o}, \frac{n_2}{n_o}, \frac{n_3}{n_o}, ....\right] , \quad (16)$$

and the constants

$$R_1 = \frac{n_i}{n_o} = \sum_{j=1}^{j_{max}} \frac{P_j}{j} . \quad (17)$$

and

$$R_2 = \frac{n_e}{n_o} = \sum_{j=1}^{j_{max}} j\, P_j \quad (18)$$

from which

$$Z = R_2/R_1 \quad (19)$$

and

$$\eta_i = (1 + 1/R_1)^{-1} . \quad (20)$$

can be computed, as well as $n_e = R_2\, [(kT_e/p)(1+R_1+R_2)]^{-1}$ (21)

for a new iteration in Eq. (15).

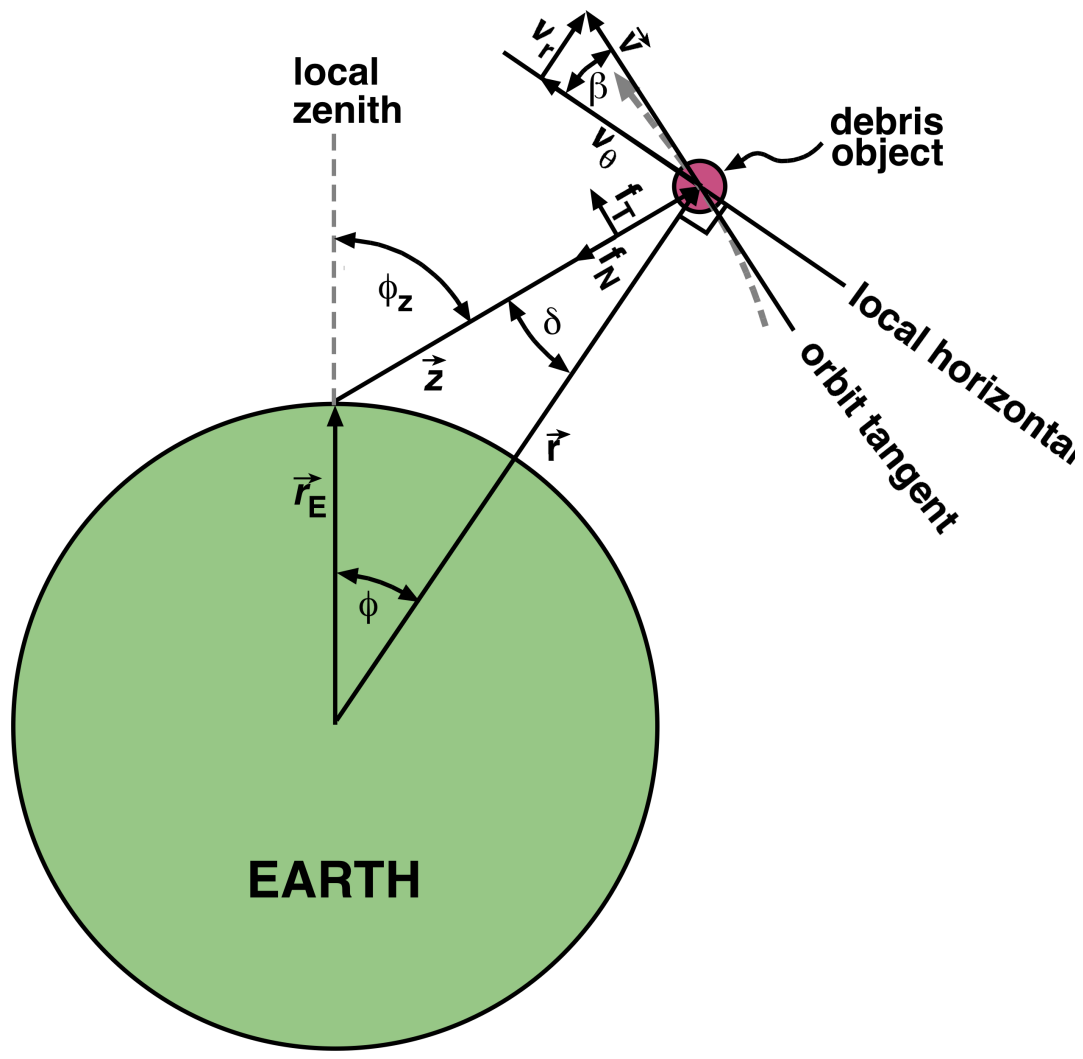

Figure 1. Geometry for orbit discussion

## Laser Orbit Modification

Figure 1 shows the geometrical variables for analyzing laser orbit modification. Where the zenith angle $\phi_z = \phi - \delta$, $\delta = -\sin^{-1}(r_E \sin\phi/z)$, and $\beta = \tan^{-1}(v_r / v_\phi)$, range to the target is obtained from

$$z^2 = r^2 + r_E^2 - 2\, r\, r_E \cos\phi \quad (22)$$

With $\bar{i}_N \bullet \bar{i}_z = -\cos(\beta - \delta) = -\cos\xi$ and $\bar{i}_T \bullet \bar{i}_z = -\sin(\beta - \delta) = \sin\xi$, and with the Hamiltonian $(E + V)$, expressed in unit mass variables,

$$E = \frac{(v_r^2 + v_\theta^2)}{2} \text{ and} \quad (23)$$

$$V = - GM/r. \quad (24)$$

The eccentricity $$e = \frac{r_a - r_p}{r_a + r_p},\tag{25}$$

where $r_a$ and $r_p$ are the apogee and perigee orbit radii. In the plane of motion, the orbit is

described by $$r(\phi) = [\frac{r_p(1+e)}{1+e\cos(\phi+\phi_o)}]\tag{26}$$

a definition which means perigee is at $\phi=\phi_o$.

Where $r_p$ is the perigee geocentric radius, and the semi-major axis $a = r_p/(1-e)$, $\ell$ is the angular momentum per unit mass, MG is the Earth's gravitational constant and the quantity

$$q = a(1-e^2) = \ell^2/\mathrm{MG},\tag{27}$$

the tangential and radial velocity components are

$$v_\phi = \sqrt{\frac{MG}{q}}[1+e\cos(\phi+\phi_o)] \quad \text{and}\tag{28}$$

$$v_r = \sqrt{\frac{MG}{q}}[e\sin(\phi+\phi_o)].\tag{29}$$

The total velocity is obtained from $v^2 = v_r^2 + v_\phi^2 = MG(\frac{2}{r} - \frac{1}{a})$. (30)

For externally perturbed orbits, we have

$$\Delta a = \frac{GM}{2H^2}\Delta H,\tag{31}$$

and $$\Delta v_r = -\Delta J_N = +\Delta J\cos\xi\tag{32}$$

$$\Delta v_\theta = +\Delta J_T = +\Delta J\sin\xi\tag{33}$$

where $\xi=\beta-\delta$. Also, $\Delta q = 2r\sqrt{p/MG}[\Delta J_T\cos\beta + \Delta J_N\sin\beta]$, (34)

or, in a more useful form, $\Delta q = \frac{2r}{v}[\Delta J_T(1+e\cos(\phi+\phi_o)) + \Delta J_N e\sin(\phi+\phi_o)]$ (35)

In Eq. (35), $\Delta J_T$ and $\Delta J_N$ are, respectively, the components of $\Delta\bar{J}$ along the orbit tangent, and along the inward normal to the orbit in the orbital plane. This equation makes the point that $\Delta J_N$ also has a major effect on the orbit, not $\Delta J_T$ alone as one might intuitively think. However, when $(\phi+\phi_o) = 0$ [perigee at zenith], Eq. (35) shows $\Delta J_N$ has no effect. We can understand that by writing $\Delta H = v_r\Delta v_r + v_\theta\Delta v_\theta$ and noting that $v_r=0$ at perigee, so that even a large $\Delta v_r$ can have no significant effect. The effect of pushing directly upward is to instantaneously tilt the velocity vector upward, so that the orbit can change later. In the majority of cases, the perigee or apogee will not be directly overhead, and calculations show we can drop perigee by pushing directly upward on the object.

Now, $$\Delta H = v_r \Delta v_r + v_\phi \Delta v_\phi , \tag{36}$$

$$\Delta v'^2 = v'^2 - v^2 = 2\Delta H , \tag{37}$$

But, since $\Delta q = (1 - e^2)\Delta a - 2ae\Delta e$, we can write (38)

giving $$\Delta e = \frac{[(1 - e^2)\Delta a - \Delta q]}{2ae} \tag{39}$$

From which, $$\Delta r_p = (1-e)\Delta a - a\Delta e \tag{40}$$

and $$\Delta r_a = (1+e)\Delta a + a\Delta e \tag{41}$$

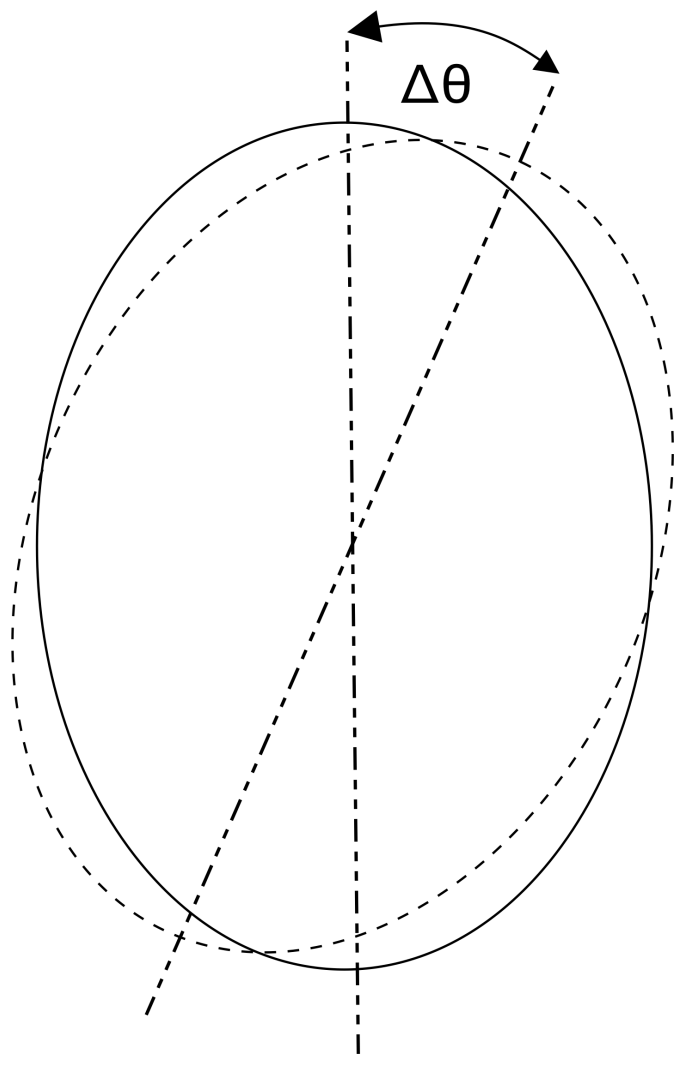

**Figure 2.** Indicating apsidal shift

which are the desired quantities. For e=0, Eq. (39) gives correct results in the limit $e \to 0$. This procedure has the advantage of being developed from first principles rather than involving intermediate relationships.

Next, we have the problem of calculating the rotation angle of the semi-major axis of the ellipse due to our actions. If it's too large, might unintentionally raise something we earlier lowered. Axis rotation can be computed. We use $\Delta\beta$ after the interaction, and $d\beta/d\phi$ for the original ellipse, to find $\Delta\theta$.

where $$\frac{d\beta}{d\phi} = \frac{d\beta}{dr}\frac{dr}{d\phi} \tag{42}$$

Since $\frac{dr}{d\phi} = -(er^2/p)\sin(\phi+\phi_o)$, with $m = (2r - r^2/a)$ and $p = a(1-e^2)$,

$$\frac{d\beta}{d\phi} = \frac{-(1 - r/a)er^2}{[1 - p/m]^{1/2} m^{3/2} \sqrt{p}} \sin(\phi + \phi_o) \tag{43}$$

and $$\Delta\theta = -\frac{\Delta\beta}{d\beta / d\phi} . \tag{44}$$

is easily calculated. For small debris, which can be re-entered in a single pass, this apsidal shift is irrelevant. For large debris, it must be taken into account when the object is re-engaged.

## Optical Constraints from the Atmosphere and Target Physics

The most complex part of a Laser ODR design is to simultaneously satisfy the constraints that arise from diffraction, nonlinear optical effects in the atmosphere and target physics.

Beam fluence in the atmosphere is constrained above and below. Where $z$ is target range, $\lambda$ is wavelength, $D_{\text{eff}}$ is launching aperture diameter, and the quantity

$$\zeta = \frac{az\sqrt{\lambda}}{D_{eff}^2} \quad (45)$$

incorporates the effects of diffraction, a minimum fluence in the atmosphere

$$\frac{\Phi_b}{\lambda} \geq \frac{\beta\zeta^2\sqrt{\tau}}{T} \quad (46)$$

is required to ignite a plasma on the target. We assume a beam quality factor of 2 and an index 6 hypergaussian radial intensity profile, which together give a = 1.7 in Eq. (45), so a typical value of $\zeta$ is 75. In Eq. (7), $T$ is atmospheric transmission, which we take to be 85%. The upper limit fluence is set by nonlinear optical (NLO) effects including (for short pulses) phase distortions due to nonlinear index ($n_2$) and stimulated rotational Raman scattering (SRS) and stimulated thermal Rayleigh scattering (STRS). For pulses 100ns≤ $\tau$ ≤1ms, the NLO effects limit amounts to $\Phi_b/\lambda \leq$ 3E10 $\tau$ $\text{Jm}^{-2}\mu\text{m}^{-1}$. For shorter pulses, this linear dependence starts to saturate, settling at $\Phi_b/\lambda \leq 100$ J $\text{m}^{-2}\mu\text{m}^{-1}$ at 100ps[l]. We can obtain solutions to these requirements graphically.

## Target Shape Effects

To draw attention to the variety of debris shapes and materials, the ORION project study[li] described five representative compositional classes: aluminum, steel, sodium-potassium spheres, carbon phenolic, and metal-coated plastic insulation. Only a fraction of these have spherical symmetry. The existence of irregularly shaped space debris brings a degree of randomness into the problem of calculating post-engagement orbital modifications: that associated with the distribution of object shapes, and that associated with orientation. Given the desire to reduce perigee, it is of interest to characterize the range of possible orbital outcomes of laser engagements with non-spherical targets (Figure 3)[lii].

In general, the impulse and laser propagation vectors are not parallel. Since ablation will be parallel to the local normal, and the impulse is directed opposite to the net ablation vector, we can write

$$m\Delta\vec{v} = -C_m \Phi_L \sum_{\alpha} A_\alpha \left|\hat{k} \bullet \hat{n}_\alpha\right| \hat{n}_\alpha \qquad (47)$$

summing over all illuminated surface elements $A_\alpha$, and the laser fluence is given by $\vec{\Phi}_L = \Phi_L\, \hat{k}$. For "smooth" objects, the sum goes over to an integral over the illuminated portion of the surface.

For illustration, we choose the simple case of a plate of mass *m*, in a low elliptical orbit, with eccentricity given by

$$\varepsilon = \left(1 + \frac{2EL^2}{G^2 M^2 m^3}\right)^{1/2} \qquad (48)$$

where $E$ is the total orbital energy, $L^2$ is the square of the orbital angular momentum, $G$ is the gravitational constant, and $M$ is Earth's mass. After engagement, a new orbit is determined from changes to $E$ and $L^2$. If the instantaneous distance from Earth's center, orbital velocity, and azimuthal velocity are denoted $r$, $\vec{v}$, and $v_\phi$, respectively, then $\Delta E$ and $\Delta L^2$ are given in terms of the velocity change by

$$\Delta E = m\vec{v} \bullet \Delta\vec{v} + \frac{1}{2} m\left|\Delta\vec{v}\right|^2$$

$$\Delta L^2 = 2m^2 r^2 v_\phi\, \Delta\vec{v} \bullet \hat{\phi} + m^2 r^2 \left(\Delta\vec{v} \bullet \hat{\phi}\right)^2 \qquad (49)$$

The quantity of primary interest is the perigee, which is

$$r_p = \frac{l^2}{GMm^2}\,\frac{1}{1+\varepsilon} \qquad (50)$$

We calculate the perigee change for a random distribution of plate orientations, and for a representative set of orbital parameters, setting $m = 1$ g for this example. The maximum laser energy on target is 10 J, which occurs when the plate is face-on to the laser position. The distribution in the perigee change at a fixed orbital angle (not shown) is weakly peaked, with substantial probability at the upper and lower bounds. Thus one can estimate the probability of achieving an undesirable result, i.e., an increased perigee, by comparing the magnitude of the upper and lower envelopes. It is also worth noting that there is a non-negligible probability of achieving a result that is *more* favorable than for the spherical case. The average perigee change for a plate is approximately 1/3 that found for a sphere, which has implications for the efficiency of targeting campaigns. Of course, this efficiency can be increased substantially by intelligently timing the laser pulse when the target and is surface orientation can be resolved.

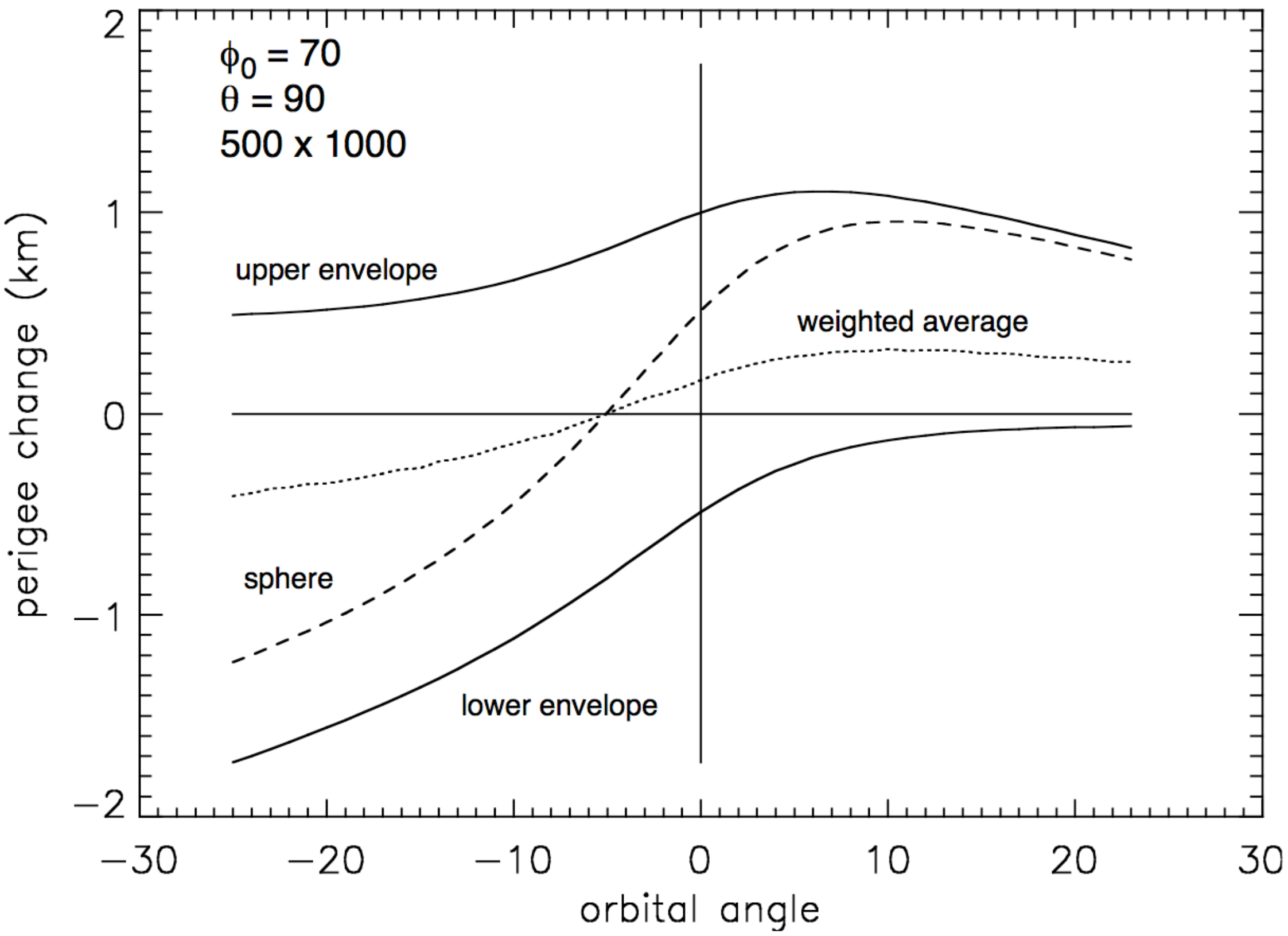

**Figure 3. Calculated perigee change.** A 1 gram plate receives a single 10 J pulse on the debris target at the indicated geocentric angle, with $C_m$ =100 μN−s/J, assuming a random distribution of plate orientations in three dimensions, perigee change plotted against orbital angle. Negative angles correspond to upstream positions relative to the laser position at ϕ=0. Horizontal extent of abscissa maps to laser horizons. Example orbit is characterized by 500 km perigee, 1000 km apogee, perigee angle ($\phi_0$) 70 degrees downstream of the laser position (descending), with an orbit intersecting laser zenith. Plotted are the best case ("lower envelope"), worst case ("upper envelope"), the weighted average (dotted), and the single-valued result for a spherical target (dashed).

It is possible for an irregularly shaped target to ablate in such a way as to create a torque about the center of mass, resulting in spin. The change in spin energy can be comparable to the change in kinetic energy. Additionally, it is known that some space debris fragments are already spinning, which means that interactions with a laser may alter the spin frequency, and may alter the orientation of the spin vector. When spin in an asymmetric debris fragment is present (or is induced), the laser/target interaction will vary from shot to shot, resulting each time in a different impulse, which leaves a complex scenario for targeting and re-acquisition. However, spin is beneficial; with several hundred engagements, the range of possible orientations becomes well sampled, and the overall effect will tend toward the mean, producing results like those in Figure 3.

## Telescope Design for Daylight Acquisition and Tracking

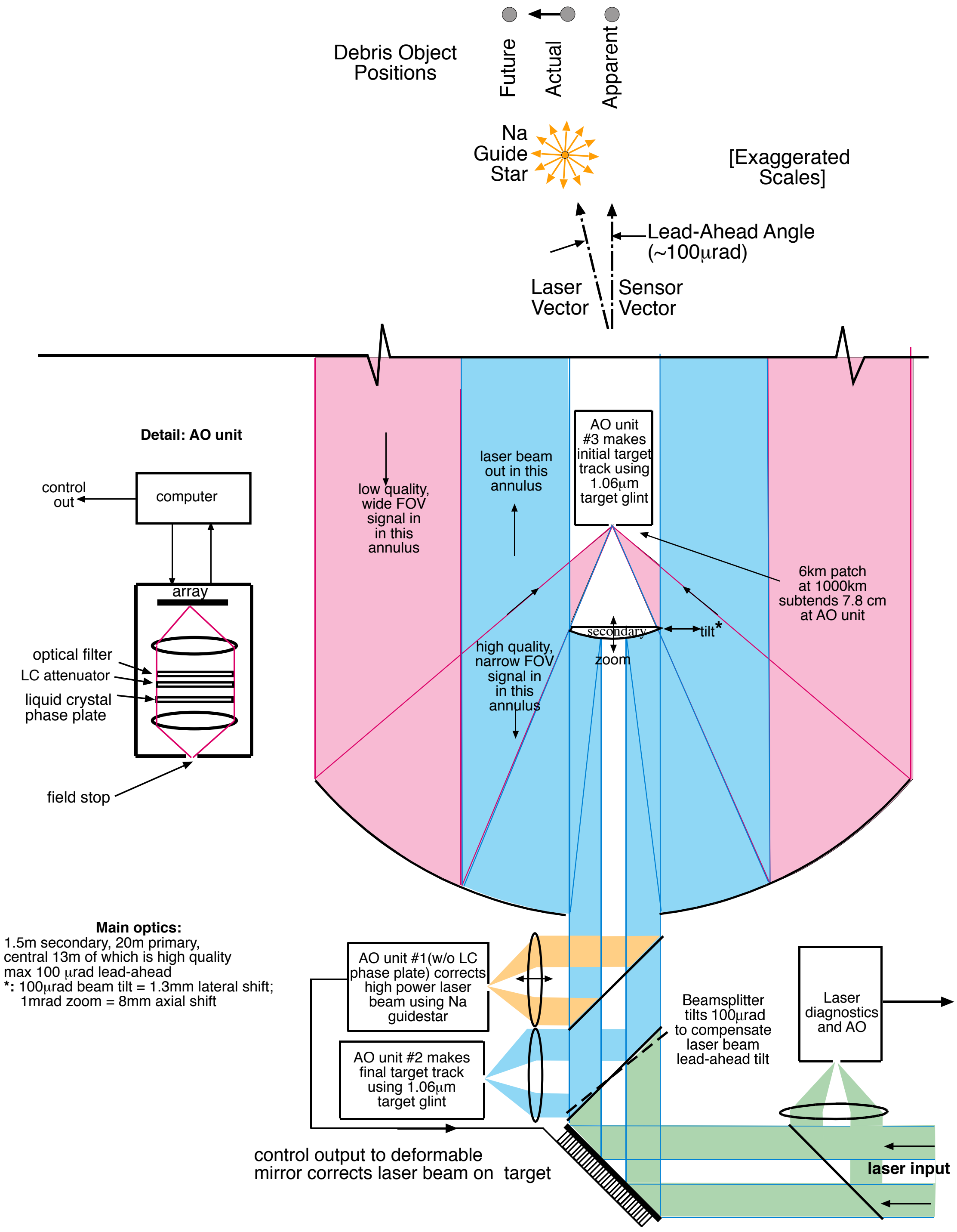

**Figure 4.** A possible telescope design discussed in the main paper.

Figure 4 shows a suggested layout for the telescope discussed in the main paper, in which standard adaptive optics are used to correct atmospheric turbulence.

## Turbulence Correction with Standard Adaptive Optics

At the bottom of Figure 4, a deformable mirror with many computer-controlled piezoelectric actuators creates a deformation in the high power laser input phase front which exactly cancels phase distortions in the atmosphere, moment by moment. Typically, a control system bandwidth of about 1kHz is required to do this. Atmospheric phase distortions are sensed by AO unit no. 1, which is pointed at the sodium guidestar which has been set up at 90km altitude by a 589.2 beacon beampath projected by the telescope (not shown). This works by exciting sodium atoms already present at that altitude, to create what is nearly a point source viewed from the ground. Knowing this, the AO system works until it sees a point source; the resulting phase shape is recorded and reversed at the deformable mirror. Of course, the laser and the guidestar are at two different wavelengths, so the computer has to attempt to calculate what the distortion should be at 1.06μm from what it knows at 589.2nm, and this is not always accurate.

Two other AO systems (no. 2 and no. 3) in the figure correct phase distortions at 1.06μm directly, using the target itself as a point source guidestar, in order to acquire the target with optimum resolution. Why then do we need the sodium guidestar? This is because AO systems no. 2 and no. 3 are pointing in a different direction from the high power laser, at the object where it appears to be. The high power laser, in contrast, has to fire into black space along a different path with different distortions at the spot where the target is predicted to be when its beam arrives. So, we need an artificial guidestar to facilitate phase correction along the high power beam path.

Note that an array of guidestars may be needed to correct for focal anisoplanatism well enough to achieve the highest possible brightness on target. Rayleigh beacons, which just use scattering from the atmosphere rather than exciting the sodium layer may also be used. These are in some ways less effective, because the effective point source is closer, but have the advantage of being at the same wavelength as the pusher laser.

## An Alternative Target Tracking Method

For active tracking, an alternate method has already been proven at the U.S. Air Force Maui Space Surveillance System (MSSS) in Hawaii. Located on the crest of Mt. Haleakala at an elevation of 3060 meters, it is also a good site for LODR because of better seeing conditions than possible at sea-level sites. One component is the Advanced Electro-Optical System 3.67-m diameter telescope at MSSS, with recoated dichroic optics and a modified coudé path. The other is the 11.2μm wavelength

"HICLASS" 900W pulsed $CO_2$ laser and its heterodyne detection system, which, together, have been shown[liii] to be able to easily track sub-cm objects at 1000km range. This performance comes about because the system is located at a cold, high altitude site, because it achieves near photon-counting performance, and because there are nine times as many photons per joule at its wavelength, compared to 1.06μm. Using this system, it should be possible to acquire and track 100 times as many targets per hour at 1000km range, these targets being twice as small, as with radar[8].

## Turbulence Correction by Brillouin-Enhanced Four-Wave Mixing

An alternative to standard adaptive optics for correcting phase distortions along the path occupied by the high power laser beam is called Brillouin-enhanced Four-wave Mixing (BEFWM). As in the standard AO technique, we depend on a few photons scattered back into the telescope when we fire the high power laser beam at the target. The debris object's reflected illumination is intercepted by the main mirror of a receiving telescope and guided to optical brightness amplifiers. After amplification, the object's image is recorded with a CCD camera. A control system turns on the brightness amplifiers and adjusts their reception frequency band. Data for the control system are provided by an illumination laser and a rangefinder that estimates moment of arrival of the scattered radiation and its frequency. The optical brightness amplifiers consist of three units: a laser amplifier, a nonlinear optical amplifier and a pump laser. To achieve the lowest level of noise in the optical brightness amplifiers, the laser amplifier gain coefficients must be about 1E4. Such a gain coefficient can be achieved in two amplification stages. When creating the laser amplifiers, it is necessary to ensure that the value of the gain coefficient is uniform over the whole field of vision. The latter, in turn, should be not less than the angle of initial illumination $\Delta\phi$ =1E-4 radians, which requires the use of optical repeaters. The space object plane is projected by a lens onto the output face of the first laser amplifier. The image is then transferred by a repeater from the output face of the first laser amplifier to the input of the second laser amplifier. Another repeater transfers images from the output face of the second laser amplifier to the nonlinear optical amplifier. The repeater is a confocal telescope.

The nonlinear optical amplifier is a Stimulated Brillouin Scattering (SBS) amplifier (a third order nonlinear medium which is active with respect to the Brillouin nonlinearity). For such a medium to amplify signal light, it should be illuminated simultaneously with the space object reflected signal and by a powerful additional laser radiation pump. Typical nonlinear medium elements are tetrachlorides such as $CCl_4$, $GeCl_4$, $SnCl_4$, or perfluorooctane. Their parameters are very similar (nonlinearity factor ~ 5E-9 cm/W, and hypersound relaxation time ~1ns). When the SBS cell is illuminated by a pulsed laser with energy 1.5 J and duration 20 ns, the SBS amplifier amplifies the space object scattered light with a gain coefficient of approximately 1E8. Therefore, the gain coefficient of light received from the space object, consisting of combined gain in laser amplifier and nonlinear optical amplifier, will be about 1E12, which is adequate to create a recordable image.

The concentration of the space object reflected laser illumination is restricted by the influence of turbulence. To improve imaging and to minimize the required level of illumination laser irradiance to make a jet on the target requires that we overcome the atmospheric turbulence to focus the beam. This is achieved through optical phase conjugation of the illumination radiation using BEFWM to reverse the beam propagation direction and phase to compensate for atmospheric distortions as the beam back propagates through the optically distorting path. If our nonlinear-optical receiver amplifies and conjugates the signal intercepted by the receiving lens, then as a result of double passage through the atmosphere turbulent distortions of the space object signal wavefront are compensated. Consequently, the conjugated signal will be concentrated on the space object to an accuracy that is determined not by the turbulent scattering angle (~10 μrad) but, instead by the resolution of the receiving aperture of the nonlinear optical amplifier (e.g., ~ 0.05 μrad for a receiving aperture of 20m).

## How BEFWM Works

A number of papers are available concerning laser propulsion[liv,lv,lvi,lvii,lviii]. System risks are low. This kind of operation on debris will not generate additional debris. Laser irradiation of large operating spacecraft will not seriously affect them, unless photo-sensitive equipment is exposed, since under the worst conditions only very small amounts of surface material are ablated.

Phase Conjugation (PC) is a non-linear optical effect that forms the same wavefront as an initial one, but which propagates exactly in the backward direction with reversed phase. A phase conjugate mirror is like a mirror reflecting incident light back towards where it came from, but it does so in a different way than a regular mirror. A regular mirror reflects incident rays in the opposite direction, so that the angle of incidence is equal to the angle of reflection (see Fig 6a of the main paper). In contrast, light from a phase conjugate mirror is always reflected exactly in the backward direction, independent of the angle of incidence.

This difference provides significant opportunities. If we place a distorting medium (e.g., a turbulent air flow) in the path of a beam of light, the rays radiated from the point-like light source are bent in random directions, and after reflection from a normal mirror, each ray of light is bent even farther causing the beam to scatter. With a phase conjugate mirror, on the other hand, each ray is reflected back in the direction it came from. This reflected conjugate wave propagates backwards through the same distorting medium, and "undoes" the distortion, causing the beam to converge to its initial point source.

Phase conjugation operates somewhat like holography, but it is a dynamic hologram whose "holographic plate" is determined by interfering waves in a nonlinear optical medium rather than etched as a static pattern on a glass plate. In our case, the physical mechanism of this process is called four-wave mixing because it is based on the interaction of laser waves and hypersonic waves. Here the interference of signal and pump laser waves creates a hypersonic grating and the second pump scattering on this grating produces a conjugated wave (Fig. 5). This interaction of laser waves with hypersound is known as stimulated Brillouin scattering and this type of four-wave mixing is called Brillouin enhanced four wave mixing (BEFWM). Figure 6 shows how a laboratory BEFWM setup works[lix].

This combination of laser amplifiers and the BEFWM PC mirror provides a unique set of capabilities that can enable and simplify the design of our debris removal system:

- The system has very high sensitivity near 4.8E-19 J per pixel (approximately two photons) which lets us minimize the laser pulse energy needed to generate a measurable scattered signal from the orbital debris.
- The system's extremely narrow frequency band corresponds to two frequency-temporal modes (input spectral bandwidth ~ 1 pm and response time of ~ 30 ns) which ensure our proposed system will reject background noise and engage extremely quickly to enable its use on hypervelocity debris.
- The system has a comparatively wide field of view that can be tailored to operational needs as part of the system design.
- The system has a high coefficient of amplification, amplifying weak signals by a factor of about 1E12.

With the concept illustrated in Figure 6, we begin by illuminating the detected orbital debris with an initial laser pulse. An input lens receives the scattered illumination from the debris to form an object image, but as this signal pulse (carrying the image) propagates through the system it is also amplified by a preliminary laser amplifier. In turn, a PC-mirror input lens focuses the object image in the PC-BEFWM mirror. This PC-mirror is a liquid cell filled with a nonlinear optical medium (typically, high-purity liquid tetrachlorides or freons) that is pumped by two pump pulses. The reflected conjugated pulse goes back to the object plane and on its way is partially reflected by the beam-splitter to the recording system (a CCD or CMOS camera), where an image of the space object is formed to allow us to identify the object as debris or not.

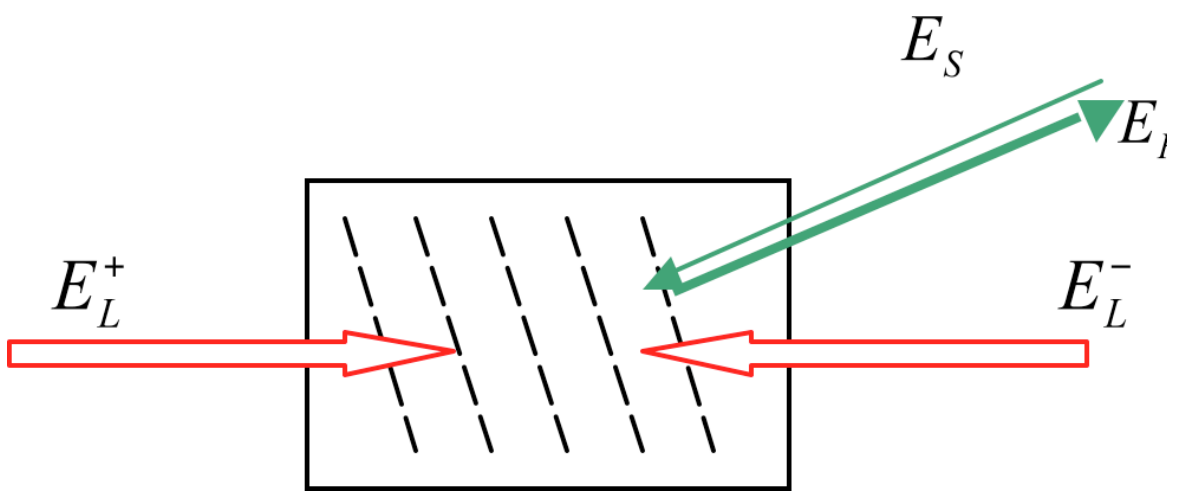

**Figure 5.** Illustrating the BEFWM process

## BEFWM system concept of operations

The system shown as Figure 6b of the main paper uses a BEFWM receiver-amplifier. First, a master oscillator MO delivers an illumination pulse of about 30ns duration at 1.06μm that is amplified by the illumination laser IL and directed to the region of space containing the debris target in a comparatively wide angle of about 100 μrad.

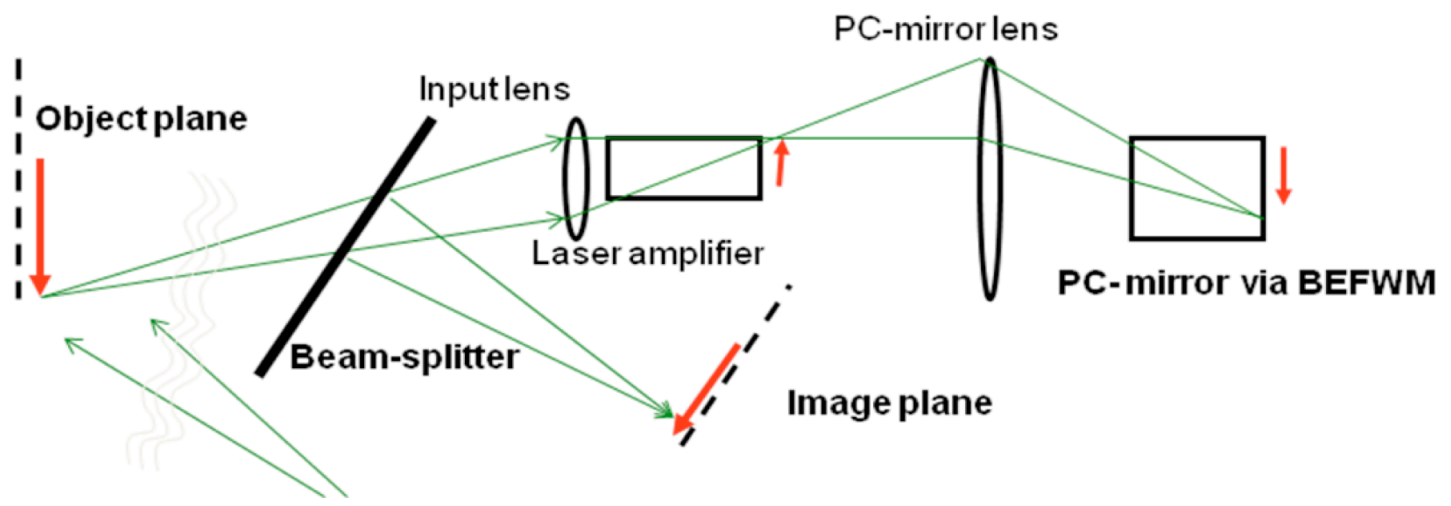

**Figure 6.** BEFWM in the laboratory

Then, part of the reflected illumination is received by an input\output (IO) mirror with clear aperture *D*, amplified and reflected by our BEFWM system and directed back toward the debris by the IO mirror. Phase conjugation provided by the BEFWM system removes the effects of atmospheric turbulence and provides perfect illumination on the debris with this second pulse, resulting in a higher quality image. The second pulse reflection from the debris may, in turn, be used to repeat this cycle to increase the lighting level, or to concentrate the laser on the debris providing a high quality glint that can be used as a target designator or to enable advanced adaptive optics methods with a guidestar maintained on the moving debris.

We assessed the concentration efficiency of the proposed system to assess the appropriate illumination pulse energy with secondary illumination using BEFWM and without it. For debris sizes on the order of 10 cm that are within a 1000 km range, we found that to image this debris without PC-adaptive optics requires high pulse energy (up to 100 kJ) for the initial illumination. However, a one-step or two-step laser energy concentration using our system provides debris imaging with reduced initial illumination pulse energy of ~ 1kJ or even 100 J.

It should be mentioned here that a primary mirror maybe of poor optical quality (reducing initial costs) because phase conjugation will correct for its distortions too. This concept should satisfy the basic requirements in terms of laser pulse delays, laser frequencies, and precise control of the pointing and signal tuning to compensate for the debris motion and Doppler shift.

Some years ago we developed a two-pulse master oscillator (MO) with controlled delay between the two pulses to compensate for the path difference between the signal and pump pulses on the way to the BEFWM-mirror, since both of the pulses must arrive there simultaneously and their frequencies should be the same.

## Long-distance open-air BEFWM demonstration

We assembled and tested our non-linear optical image amplifier scheme in outdoor experiments through turbulence[lx]. Our path length for open air experiments was short (150m), so we developed a method of controlled turbulence intensification using banks of heaters to simulate a longer path. This heating increased the structural parameter of atmospheric turbulence by a factor of 5-10, corresponding to a propagation path of several km. Our test path with air heaters and target area at the background is shown in Fig. 7 (left image). Laser energy concentration was demonstrated in the experiments. The target area is shown in the right image. We used a glass spherical reflector on a tripod to imitate a point target. In the upper right corner of Figure 7 there is a print of initial illumination of a point-like target. The second picture is a print of both pulses (the initial pulse and the pulse reflected and concentrated by BEFWM) simultaneously. Using an oscilloscope placed in the target area we verified that the second pulse with less total energy has a much higher laser energy density. To demonstrate this fact, a glass beam-splitter was placed in front of the spherical reflector to reflect part of the signal to a fast photo-diode. We achieved similar results in other experiments carried out on a 2.1km path and with pulse energy up to 100 J. We concluded that we can mitigate atmospheric turbulence to provide near diffraction limited images and focus a pulsed laser on orbital debris.

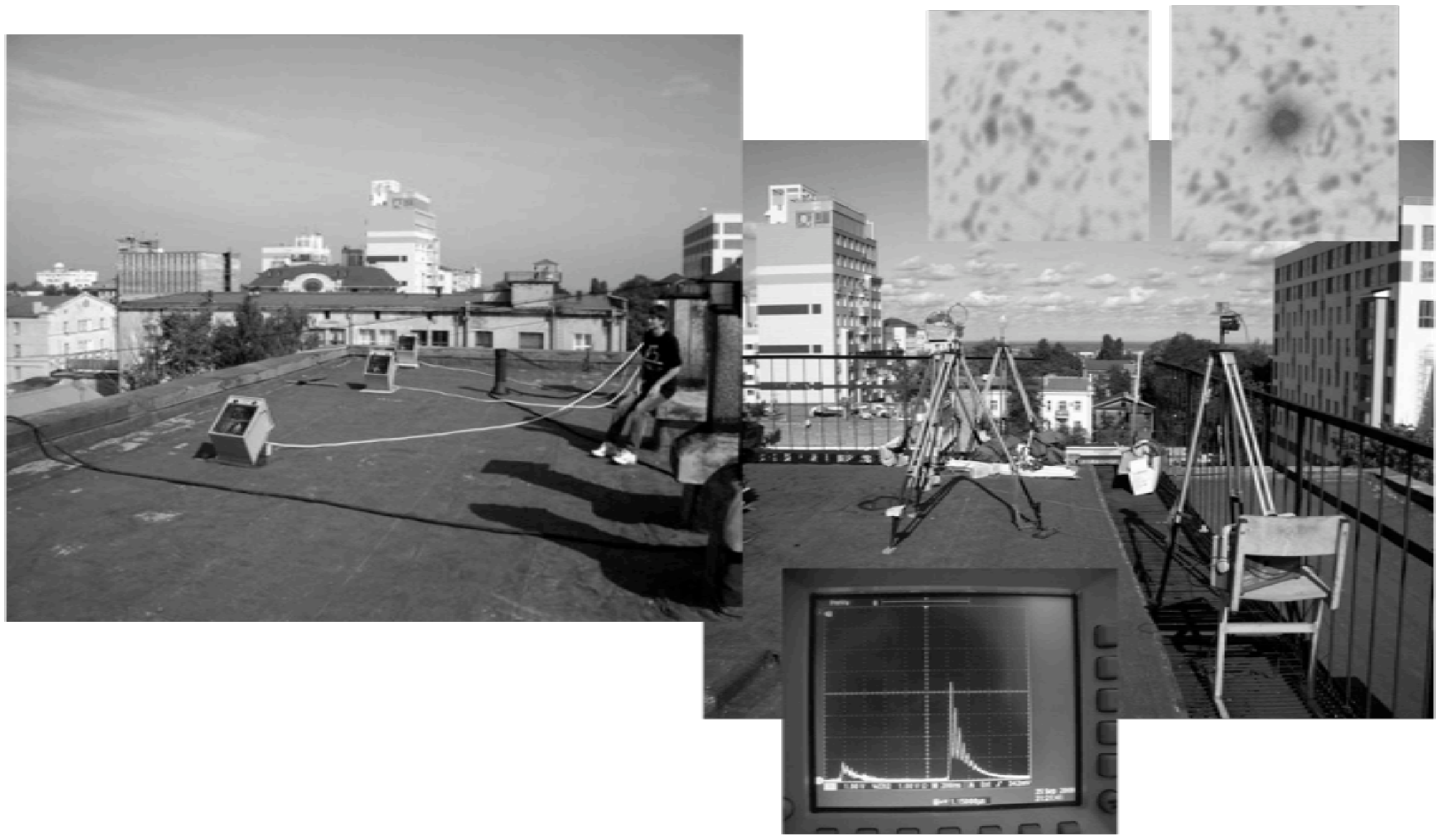

**Figure 7.** Highlights of open-air BEFWM experiments

## Choosing Targets

The LODR system target selection balances priority with routine tasking operations, which include laser and surveillance system tasking for efficient utilization of resources. Based on SOCRATES and other space surveillance and conjunction estimates already supported by the USAF, priority operations will task the LODR to deflect a potential threat to a high value asset such as Space Station by applying a Δv as small as 20 m/sec, and tasking additional surveillance for post illumination track maintenance. The routine space debris clearing will select targets with acceptable engage ability and safety. The overall concept of operations (CONOPS) is expected to consider uncertainties in target cross section, orientation and spin rate, target materials and mass, required delta V for assured re-entry and potential for fragmentation and collateral threat. Smaller debris single pass target illuminations at low laser beam elevations will be most effective by slowing the target by 50 to 200 m/sec and thus dropping its perigee for a rapid re-entry. Larger and heavier targets will require a multi-orbit plan for gradually lowering the perigee, and additional surveillance resources will be needed to maintain tracking on a perturbed orbit with potentially changed drag characteristics.. A dual site LODR would provide additional access and response capability. Figures 8 and 9 outline the concept of operations.

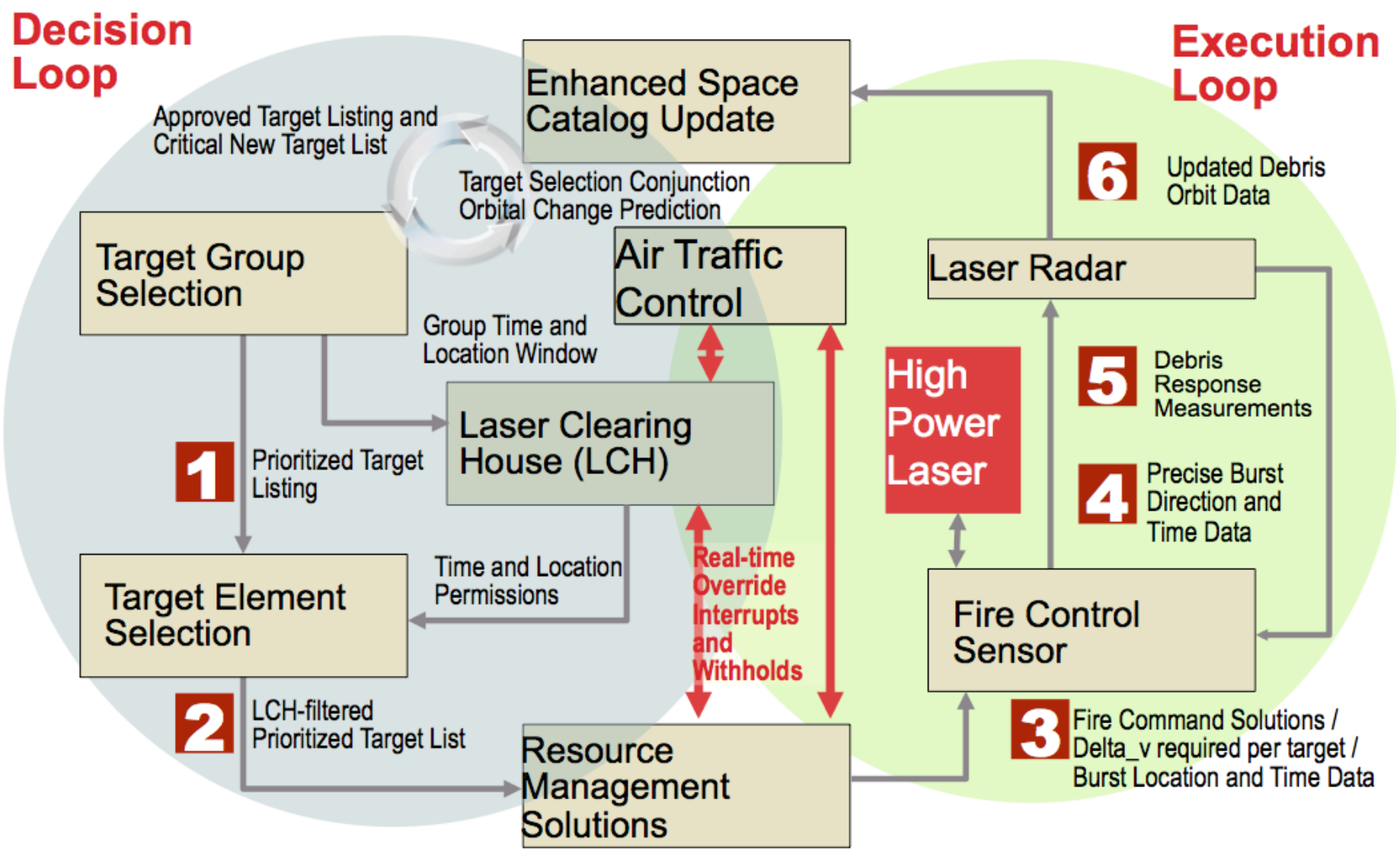

**Figure 8.** Concept of operations for a LODR removal system

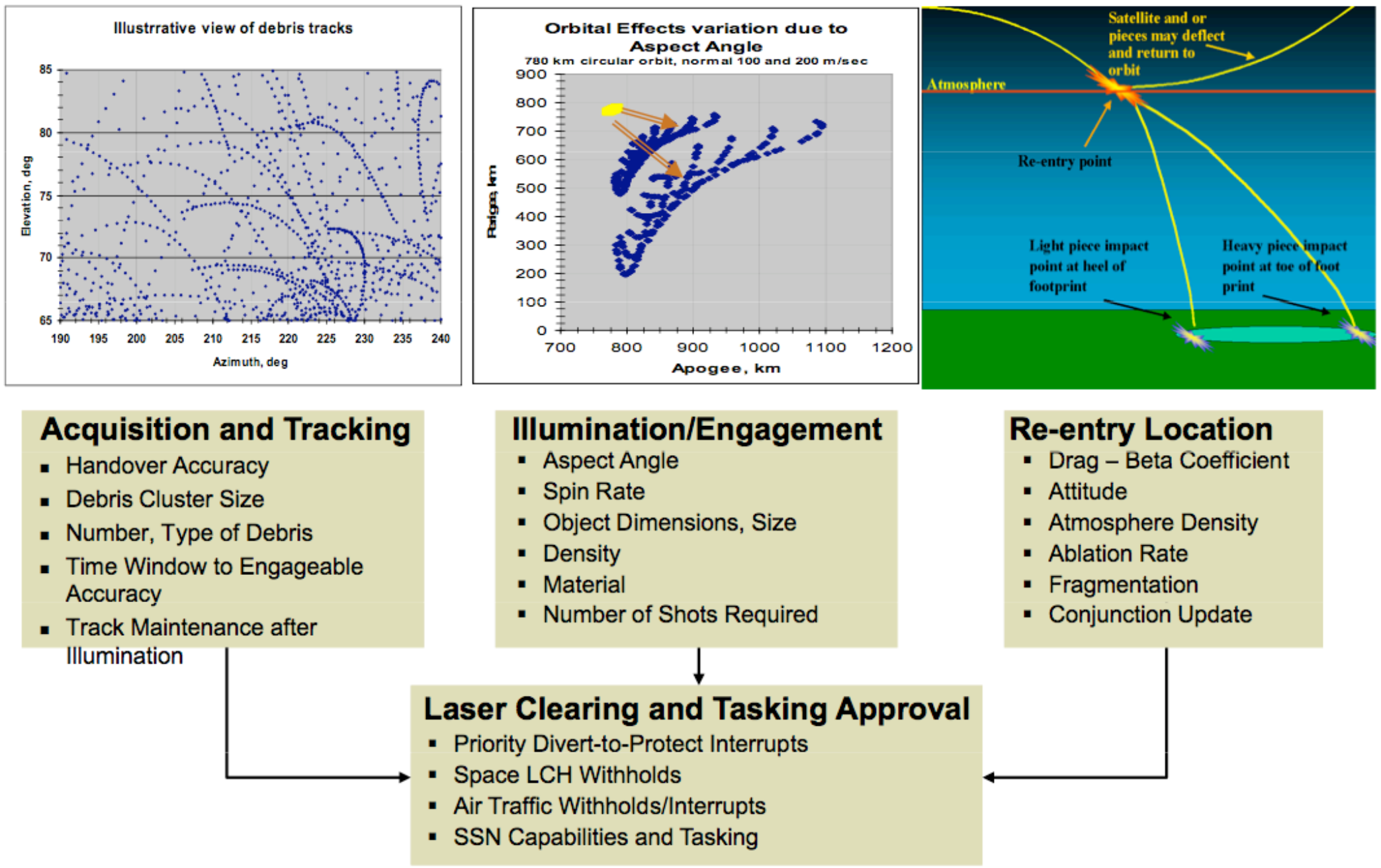

**Figure 9.** Balancing four areas of uncertainty

# References

[xviii] I. Lee, "An Investigation into Passive Collision Removal Methods," *NASA/DARPA Orbital Debris Conference… ibid.*

[xix] K. Leavitt, "NASA Ames ADR Efforts," *NASA/DARPA Orbital Debris Conference… ibid.*

[xx] D. McKnight, "Removing Orbital Debris without Going into Orbit," *NASA/DARPA Orbital Debris Conference… ibid.*

[xxi] T. Hanada, Y. Kitazawa, A. Kawabe, H. Matsumoto and S. Kawamoto, "Small and Medium Orbital Debris Removal Using Special Density Material," *NASA/DARPA Orbital Debris Conference… ibid.*

[xxii] C. Phipps, J. Reilly and J. Campbell, "Optimum Parameters for Laser Launching Objects into Low Earth Orbit," *Laser and Particle Beams* vol. 18, pp 661-695 (2000)

[xxiii] C. Phipps, "'Catcher's Mitt' as an Alternative to laser Space Debris Mitigation," *AIP Conference Proceedings* volume 1278 pp. 509-514(2010)

[xxiv] J. Mason, J. Stupl, W. Marshall and C. Levit, "Orbital Debris Collision Avoidance," arXiv:1103.1690v1 [physics.space-ph] 9 Mar 2011

[xxv] J. – C. Liou and N. L. Johnson, "Risks in Space from Orbiting Debris," *Science* vol. 311, pp. 340-341 (2006)

[xxvi] J. Sinko and C. Phipps, "Modeling $CO_2$ laser ablation impulse of polymers in vapor and plasma regimes," *Appl. Phys. Lett.* **95**, 131105 (2009)

[xxvii] C. Phipps, "An Alternate Treatment of the Vapor-Plasma Transition," *Int. J. Aerospace Innovations* **3**(1) pp. 45-50 (2011)

[xxviii] C. R. Phipps, Jr., T. P. Turner, R. F. Harrison, G. W. York, W. Z. Osborne, G. K. Anderson, X. F. Corlis, L. C. Haynes, H. S. Steele, K. C. Spicochi and T. R. King, "Impulse Coupling to Targets in Vacuum by KrF, HF and $CO_2$ Lasers" , *J. Appl. Phys.,* **64**, 1083 (1988).

[xxix] C. Phipps, M. Birkan, W. Bohn, H.-A. Eckel, H. Horisawa, T. Lippert, M. Michaelis, Y. Rezunkov, A. Sasoh, W. Schall, S. Scharring and J. Sinko, "Laser Ablation Propulsion," *J. Propulsion and Power*, **26** no. 4 pp. 609-637 (2010)

[xxx] C. R. Phipps, S. J. Thomas and D. E. Watkins, "Effect of nonlinear refraction on beam brightness in laser fusion applications," Proc. Intl. Conf. on Lasers '79, STS Press, McLean VA (1980) pp. 878-887

[xxxi] R. Egerman, S. De Smitt and D. Strafford, "Low-weight, low-cost, low-cycle time, replicated glass mirrors," *Proc. SPIE* **7739** p. 77390G (2010)

[xxxii] D. Strafford, S. DeSmitt, P. Kupinski and T. Sebring, "Development of lightweight stiff stable replicated glass mirrors for the Cornell Caltech Atacama Telescope (CCAT)," *Proc. SPIE* **6273** p 62730R (2006)

[xxxiii] http://www.eso.org/public/teles-instr/e-elt.html (2011)

[xxxiv] D. Strafford, "CCAT Panels: Corrugated mirror solution," http://www.submm.org/mtg/2006/2006-01-pasadena/11_Borosilicate_Panels.pdf (2006)

[xxxv] D. Liedahl, S. Libby, A. Rubenchik, "Momentum Transfer by Laser Ablation of Irregularly Shaped Space Debris," International Symposium on High Power Laser Ablation, Santa Fe, N.M., 2010, *AIP Conf. Proc.* **1278**, pp. 772-779 (2010)

[xxxvi] H. Klindrad, *ibid.,* p. 70